\documentclass[pdflatex,sn-basic]{sn-jnl}% Math and Physical Sciences Numbered Reference Style 
\usepackage{natbib}
\usepackage{graphicx}%
\usepackage{multirow}%
\usepackage{mathrsfs}%
\usepackage[title]{appendix}%
\usepackage[table]{xcolor}
\usepackage{tabularx}
\usepackage{textcomp}%
\usepackage{manyfoot}%
\usepackage{booktabs}%
\usepackage{algorithm}%
\usepackage{algorithmicx}%
\usepackage{algpseudocode}%
\usepackage{listings}%
\usepackage[inline]{enumitem}
\usepackage{subcaption}
\usepackage{soul}
\usepackage{geometry}
\usepackage{anyfontsize}
\usepackage{xurl}
\usepackage{lscape}
\usepackage[super]{nth}
\usepackage{tabularray}
\usepackage[most]{tcolorbox}
\usepackage{changepage}
\usepackage[group-separator={,}]{siunitx}

\makeatletter
\long\def\@makefntext#1{\leavevmode
  \@makefnmark\nobreak
  #1%
}
\makeatother

\newboolean{showcomments}
\setboolean{showcomments}{true} % comment this line to deactivate comments

\ifthenelse{\boolean{showcomments}}
{
	\newcommand{\del}[1]{\textcolor{red}{\sout{#1}}}    % please delete
}{
	\newcommand{\del}[1]{}                              % please delete
	
}

\ifthenelse{\boolean{showcomments}}{
	\newcommand{\nbc}[3]{
		{\colorbox{#3}{\bfseries\sffamily\scriptsize\textcolor{white}{#1}}}
		{\textcolor{#3}{\sf\small$\langle$\textit{#2}$\rangle$}}}
}{
	\newcommand{\nbc}[3]{}
	
}

\newtcolorbox{rqanswer}{
    enhanced,
    colback=gray!20,
}

\begin{document}

\title[Article Title]{Software Testing for Extended Reality Applications: A Systematic Mapping Study}


\author*[1]{\fnm{Ruizhen} \sur{Gu}}\email{rgu10@sheffield.ac.uk}

\author[1]{\fnm{José Miguel} \sur{Rojas}}\email{j.rojas@sheffield.ac.uk}

\author[1]{\fnm{Donghwan} \sur{Shin}}\email{d.shin@sheffield.ac.uk}

\affil[1]{\orgdiv{School of Computer Science}, \orgname{University of Sheffield}, \orgaddress{\city{Sheffield}, \country{UK}}}

% \affil[2]{\orgdiv{Department}, \orgname{Organization}, \orgaddress{\street{Street}, \city{City}, \postcode{10587}, \state{State}, \country{Country}}}

% \affil[3]{\orgdiv{Department}, \orgname{Organization}, \orgaddress{\street{Street}, \city{City}, \postcode{610101}, \state{State}, \country{Country}}}

% \newcommand\rev[1]{\textcolor{black}{#1}}

\abstract{
Extended Reality (XR) is an emerging technology spanning diverse application domains and offering immersive user experiences. However, its unique characteristics, such as six degrees of freedom interactions, present significant testing challenges distinct from traditional 2D GUI applications, demanding novel testing techniques to build high-quality XR applications.
This paper presents the first systematic mapping study on software testing for XR applications. We selected 34 studies focusing on techniques and empirical approaches in XR software testing for detailed examination. The studies are classified and reviewed to address the current research landscape, test facets, and evaluation methodologies in the XR testing domain.
Additionally, we provide a repository summarising the mapping study, including datasets and tools referenced in the selected studies, to support future research and practical applications.
Our study highlights open challenges in XR testing and proposes actionable future research directions to address the gaps and advance the field of XR software testing.
}




\maketitle

\section{Introduction} \label{sec:intro}

% background of XR
The global market for Extended Reality (XR) has grown significantly in recent years ---estimated at USD 77 bn in 2024--- and is expected to continue its rapid expansion to cross USD 3 tn by 2037\footnote{Market forecast available at: \url{https://www.researchnester.com/reports/extended-reality-market/4863}}, reflecting the increasing adoption and technological maturation of XR across multiple sectors.
% \rev{
The industry continues to evolve with major technology companies investing heavily in this space. In late 2024, Google announced Android XR, a dedicated XR operating system built for next-generation computing experiences\footnote{\url{https://blog.google/products/android/android-xr/}}. The platform is developed in collaboration with Samsung for their forthcoming headset (expected in 2025), which might represent a significant shift in the XR landscape, further accelerating mainstream adoption.
% }
%
While entertainment--particularly video games--remains the most popular application domain for XR technologies~\citep{rodriguez2017EmpiricalStudyOpen}, various other fields have also benefited from its rapid development, including education~\citep{kavanagh2017SystematicReviewVirtual}, engineering~\citep{Tadeja2020AeroVR}, military~\citep{lele2013VirtualRealityIts}, and medicine~\citep{kim2017VirtualRealityAugmented}. This broad spectrum of applications underscores the transformative potential of XR technologies beyond consumer entertainment.

XR is an umbrella term encompassing Augmented, Mixed and Virtual Reality (resp. AR, MR and VR). XR applications (hereafter, XR apps) are software programs designed to run on XR-compatible devices. These apps typically feature virtually organised spaces populated with virtual objects and interactive elements, allowing users to explore scenes and engage with digital content.
For instance, \emph{Pok\'emon Go}\footnote{\url{https://pokemongolive.com}}, a phenomenal AR mobile game, utilises GPS and cameras of mobile devices to overlay virtual content onto real-world locations.
% \ds{I think we should explain somewhere that XR includes VR and AR.}
%
More immersive experiences are offered through head-mounted displays (HMDs), such as VR headsets (e.g., PlayStation VR2\footnote{\url{https://playstation.com/ps-vr2}}) and AR headsets (e.g., Apple Vision Pro\footnote{\url{https://apple.com/apple-vision-pro}} and Meta Quest 3\footnote{\url{https://meta.com/gb/quest/quest-3}\label{footnote:MetaQuest}}).

% software testing for XR
As XR apps become increasingly prevalent across diverse and critical domains and multiple platforms and devices, their development and testing have grown significantly more complex~\citep{andradeUnderstandingVRSoftware2020-PS29}.
XR apps possess unique characteristics that distinguish them from traditional apps, such as mobile 2D apps. These include real-time responsiveness and complex interactions, enabling users to \emph{select} and \emph{manipulate} virtual objects or \emph{navigate} through virtual environments~\citep{VR/AR}.
These differences pose unique challenges for software testing. For example, in the context of generating test sequences, Android apps have finite interaction paths when navigating between different \emph{activities} (i.e., individual screens of an app)~\citep{su2017GuidedStochasticModelbased}. In contrast, XR apps involve virtually infinite interaction possibilities; even a simple task, such as moving towards and interacting with a virtual object, requires accounting for countless variations in interaction sequences~\citep{deandrade2023ExploitingDeepReinforcement-PS17}. These complexities necessitate advanced software testing methods to ensure that XR apps operate reliably and meet user expectations.

% existing testing approaches and limitations
% \rev{
Many XR platforms now include simulation capabilities that allow developers to test and debug apps without requiring physical headset usage, such as Meta XR Simulator\footnote{\url{https://developers.meta.com/horizon/documentation/unity/xrsim-intro}} and Unity XR Device Simulator\footnote{\url{https://docs.unity3d.com/Packages/com.unity.xr.interaction.toolkit@3.0/manual/xr-device-simulator-overview.html}}. 
For instance, Meta XR Simulator supports Meta Quest app development by enabling keyboard, mouse, or game controller simulation of XR interactions. The simulator also features a valuable \emph{record and replay} function that captures input sequence and verifies consistent behaviour across executions. While record-and-replay is a common testing approach for GUI apps that simplifies the automation of complex usage scenarios~\citep{Hu2015RecordReplay, Modarressi2024CaptureAR}, it has limitations, such as poor maintainability, where the captured test frequently breaks when the app's UI changes, requiring substantial manual updates to remain effective~\citep{Lam2017RecordReplay}.
% }

% Why need this study
Unlike traditional software, which benefits from well-established surveys covering various testing practices~\citep{zein2016SystematicMappingStudy, GAROUSI20131374}, to the best of our knowledge, there are currently no available comprehensive systematic review studies dedicated to XR software testing. Critical aspects, such as testing practices, tools, frameworks, and general testing guidelines, remain largely unexplored.
% \ds{perhaps we could mention well-cited existing testing surveys? not necessary though}
% \ds{Can we say some general limitations of the lack of a systematic literature survey or mapping study in a field?}

To address this gap, we present a systematic mapping study on XR software testing. Systematic mapping is a methodology designed to survey the literature, provide a comprehensive overview of a topic, identify research gaps, and offer insights into future research directions.
By carefully following the guidelines proposed by~\citet{petersen2015GuidelinesConductingSystematic}, we selected a total of 34 primary studies as the subjects for this mapping (see Appendix~\ref{appendix:PS} for the details of the studies). 
%
% A new classification scheme was built to structure the research area of XR software testing\ds{Is the new classification scheme one of the main contributions? Since it comes first, it looks like it's the main contribution.}. 
%
% Through a rigorous discussion among the authors\ds{need to rephrase this sentence before, but we needed a short ``how'' explanation here}, we classified the studies into seven specific research topics:
% %
% \begin{enumerate*}[label=(\arabic*)]
%     \item XR-specific testing,
%     \item scene testing,
%     \item automated testing,
%     \item usability testing,
%     \item security testing,
%     \item test automation, and
%     \item empirical study.
% \end{enumerate*}
% \ds{how about commenting out the above like this?}
We systematically classified and extracted data from these studies to investigate the current research status in XR software testing, explore key testing facets (e.g., activities and objectives), and examine the evaluation methodologies used.
% \ds{are they consistent with our research questions? Just to check}
%
To facilitate future research, we compile and present the tools and datasets used in the studies.
Finally, we identify the limitations and challenges in XR software testing and highlight potential avenues for advancing the field.

The main contributions of this systematic mapping study are as follows:
\begin{itemize}
    \item We provide an in-depth survey of the current software testing methods for XR apps, shedding light on the state-of-the-art in this emerging domain. The data extraction template used to derive these findings is included as part of the study.
    \item We compile a repository of existing tools and datasets used in XR software testing to support future research in the field. The repository, along with the data extraction results, is publicly available at: \url{sites.google.com/view/xr-testing}.
    \item We identify critical challenges in XR software testing and outline potential research directions to address these challenges and advance the field.
\end{itemize}

This paper is structured as follows.
Section~\ref{sec:background} presents the background of this work, key definitions and a primer on XR user interaction.
Section~\ref{sec:related_work} discusses the motivation behind this work and summarises relevant related studies.
Section~\ref{sec:methodology} details our methodology, including the process for searching and selecting relevant literature.
Section~\ref{sec:results} presents the results of our study and answers to our research questions. %, including the classification of research and data extraction from selected studies.
Section~\ref{sec:discussion} discusses the findings and explores their implications for the field of XR software testing.
Section~\ref{sec:conclusion} summarises the key contributions and concludes the paper.

\section{Background} \label{sec:background}

This section provides background on Extended Reality (XR), covering key concepts and terminology across various immersive technologies that form the foundation of this mapping study. Table~\ref{tab:abbreviations} lists abbreviations frequently used throughout this paper. 
We introduce the nature of XR applications and their user interaction models, followed by relevant software testing concepts--particularly focusing on automated testing, test automation, and GUI testing approaches.

\begin{table}[t]
    \rowcolors{1}{white}{gray!20}
    \renewcommand{\arraystretch}{1.3}
    \centering
    \caption{Summary of Frequent Abbreviations}
    % \rev{
    \begin{tabularx}{\linewidth}{cX} \toprule
        \textbf{Abbrev.} & \textbf{Definition} \\ \midrule
        XR & Extended reality, an umbrella term encompassing augmented (AR), mixed (MR), and virtual (VR) reality technologies. \\ \midrule
        VR & Virtual reality, a computer-generated simulation that immerses users in a virtual environment using 3D displays and motion tracking. \\ \midrule
        AR & Augmented reality, a technology overlaying digital content onto the real-world visual environment. \\ \midrule
        MR & Mixed reality, a technology blending real and virtual worlds, allowing physical and digital objects to interact. \\ \midrule
        DOF & Degree of freedom, the number of independent ways an object can move or rotate in three-dimensional space. \\ \midrule
        HMD & Head-mounted display, a wearable display device positioned in front of the user's eyes to provide immersive visual experiences. \\
        \bottomrule
    \end{tabularx}
    % }
    \label{tab:abbreviations}
\end{table}

\subsection{Extended Reality}

Over recent years, the development of virtual technologies, such as VR and AR, has grown rapidly. 
These advancements allow users to immersively interact with virtual objects and virtual environments with specific devices, such as HMDs and controllers.

\begin{figure*}[t]
    \centering
    \includegraphics[width=0.85\linewidth]{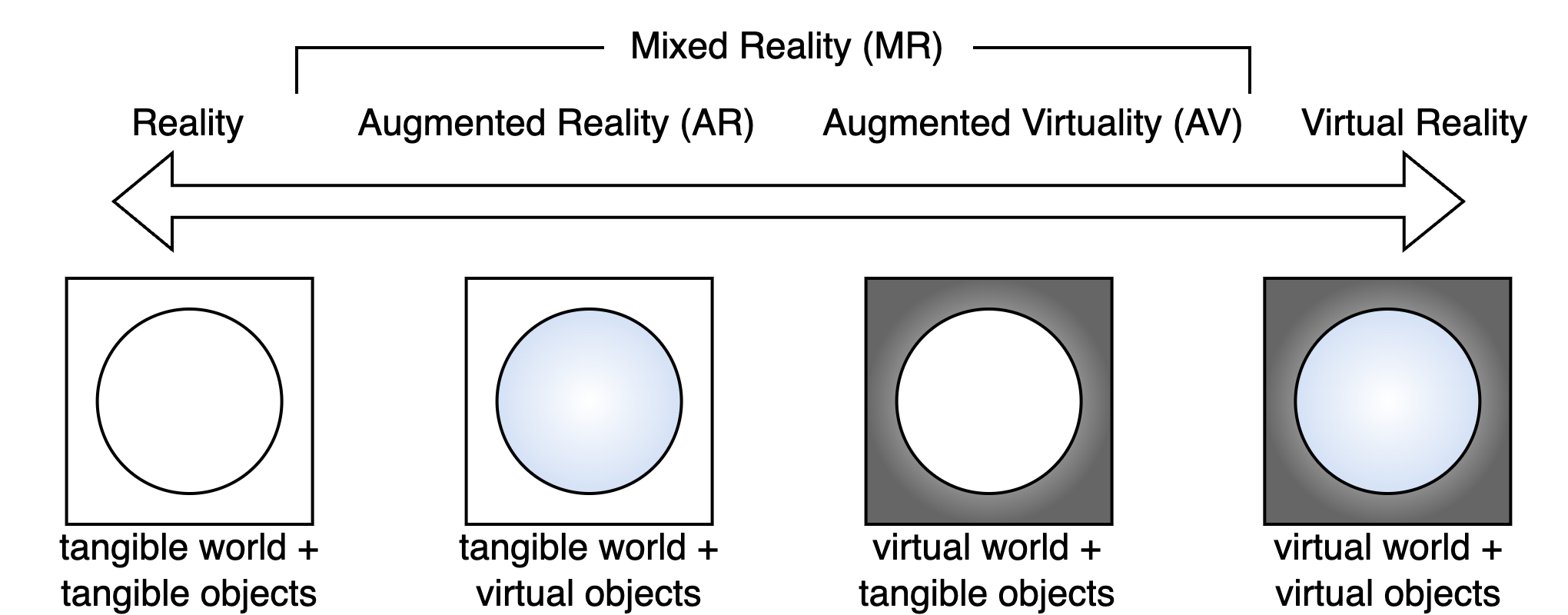}
    \caption{Reality-Virtuality Continuum}
    \label{fig:RV_continuum}
\end{figure*}

Figure~\ref{fig:RV_continuum} illustrates the differences between XR technologies using the reality-virtuality continuum introduced by \citet{milgram1994AugmentedRealityClass}. %, which is widely accepted in the research community.
The continuum spans from fully real environments (reality, on the left) to entirely virtual ones (virtual reality, on the right). The proportion of real versus virtual elements shifts along the continuum: reality diminishes while virtuality increases, and vice versa. AR, MR, and VR represent distinct forms of XR across this spectrum.

\begin{figure}
    \centering
    \begin{subfigure}[b]{.32\linewidth}
        \centering
        \includegraphics[width=1\linewidth]{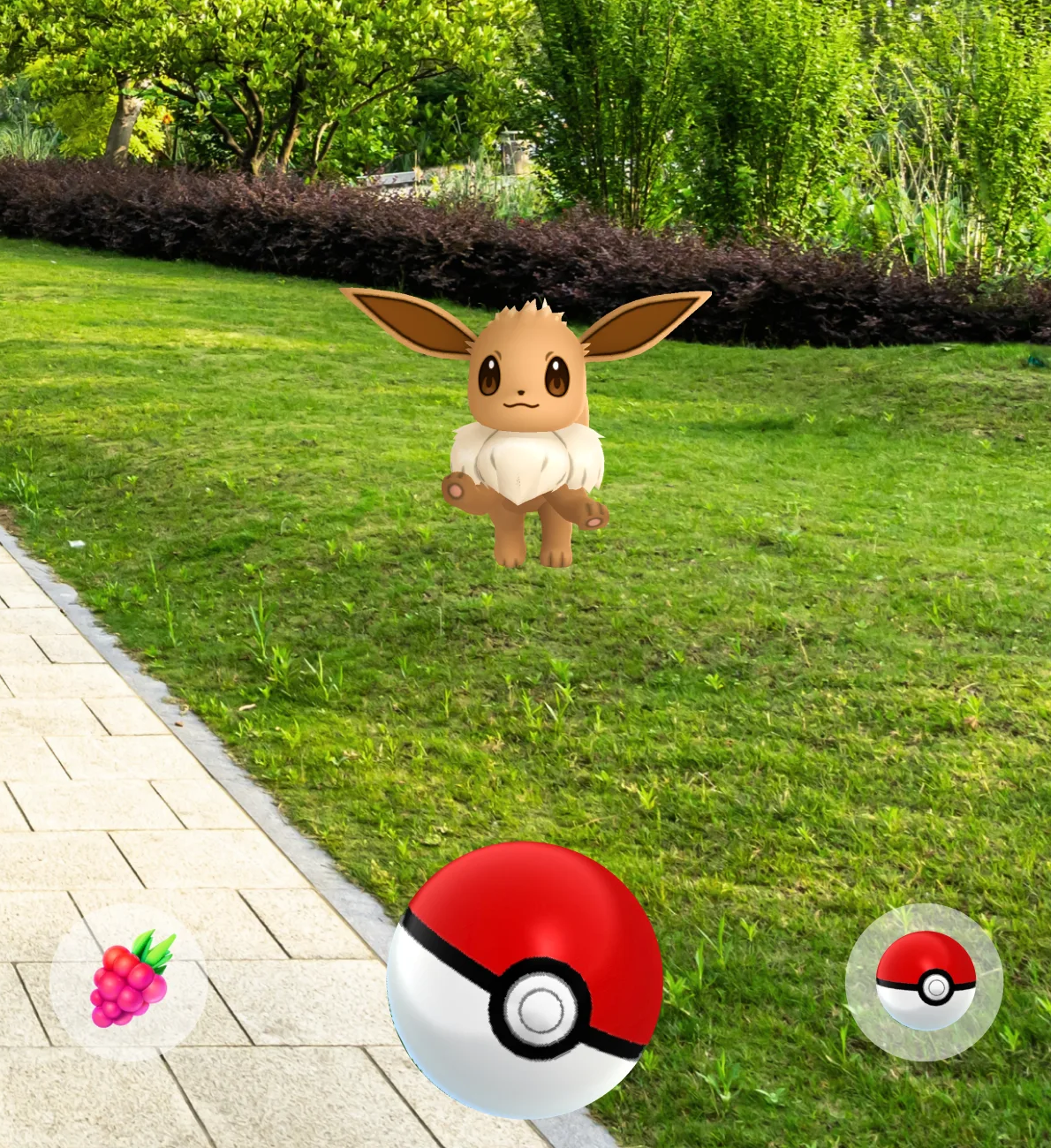}
        \caption{AR Example}
        \label{fig:ar_scene}
    \end{subfigure}
    \hfill
    \begin{subfigure}[b]{.63\linewidth}
        \centering
        \includegraphics[width=1\linewidth]{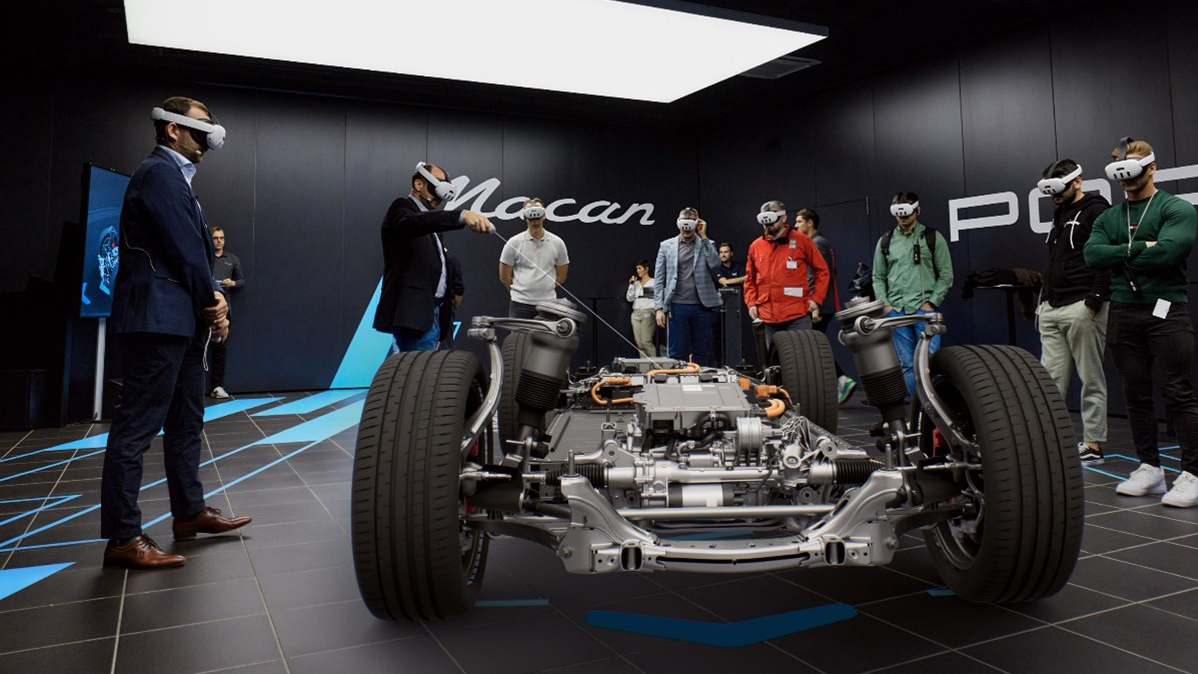}
        \caption{MR Example}
        \label{fig:mr_scene}
    \end{subfigure}
    \hfill
    \begin{subfigure}[b]{.6\linewidth}
        \centering
        \includegraphics[width=1\linewidth]{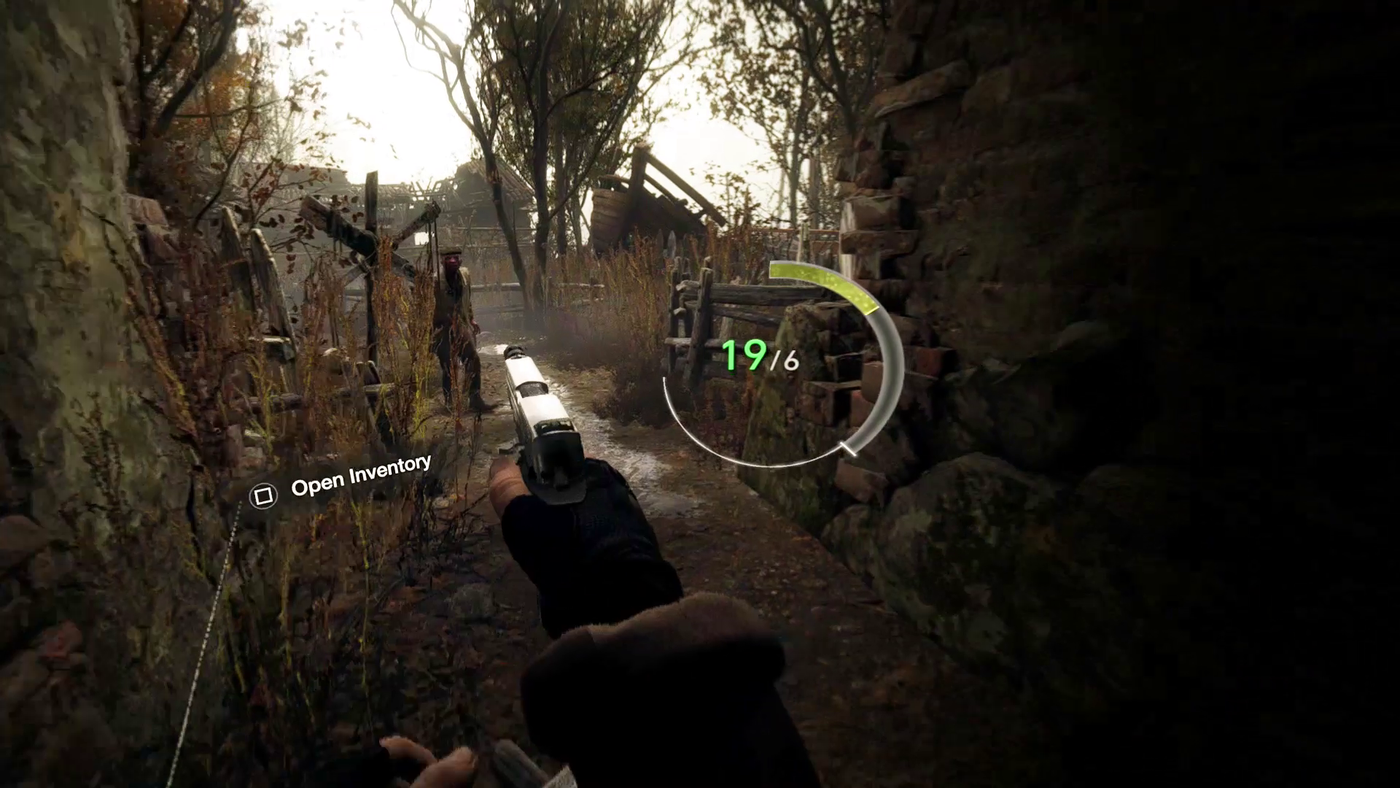}
        \caption{VR Example}
        \label{fig:vr_scene}
    \end{subfigure}
    \caption{Examples of AR and VR Scenes}
    \label{fig:vr_ar}
\end{figure}

% \subsubsection{Augmented Reality}

\textbf{Augmented Reality (AR)} is positioned near the reality end of the continuum; it overlays virtual objects onto the real world in real time, allowing users to interact with both. A prominent example is the mobile game \textit{Pok\'emon Go}~(Figure~\ref{fig:ar_scene}), where virtual creatures and widgets are superimposed onto real-world environments.

\textbf{Mixed Reality (MR)} bridges AR and VR by blending real and virtual environments, enabling real-time interaction between physical and digital elements. Virtual objects in MR behave as if they existed in the real world, offering enhanced functionality and immersion. For instance, car designers can use MR to manipulate 3D models of car components, refining designs with seamless interaction between real and virtual elements~(Figure~\ref{fig:mr_scene}\footnote{\url{https://newsroom.porsche.com/en/2024/innovation/porsche-mixed-reality-workshop-augmented-reality-34998.html}}). This capability goes beyond enhancing real-world experiences with added information by allowing deeper integration and interaction between realms.

As XR technologies continue to evolve, the distinction between AR and MR has become blurred, with the terms AR and MR often used interchangeably in both industrial and academic contexts. ~\citep{VR/AR}.
For clarity throughout this paper, we maintain the distinction between these two technologies based on the definitions provided above, with AR focusing on overlaying information and MR enabling deeper integration between real and virtual elements.

\textbf{Virtual Reality (VR)} is located at the virtuality end of the continuum, VR immerses users entirely in a digital environment, blocking out the real world. For instance, Figure~\ref{fig:vr_scene} shows \emph{Resident Evil 4 VR Mode}\footnote{\url{https://store.playstation.com/en-gb/product/EP0102-PPSA07412_00-RE4RDLC000000028}}, a VR video game, where players perform actions like shooting and reloading within a fully virtual setting.

\subsubsection{XR Applications} \label{sec:xr_app}

Most XR apps are developed using 3D engines and platforms, like Unity\footnote{\url{https://unity.com/}}and Unreal Engine\footnote{\url{https://www.unrealengine.com/}}~\citep{roberts2023ARVRTechnology}, which support deployment on various platforms, such as Android, iOS and the web~\citep{scheibmeir2019QualityModelTesting-PS20, qiao2019WebARPromising}.
%
% These tools support the deployment of XR applications on various platforms, with mobile platforms like Android and iOS being particularly popular for augmented reality applications due to their compatibility with various development tools such as ARCore and ARKit~\citep{scheibmeir2019QualityModelTesting-PS20}.
%
% Beyond mobile devices, the web-based XR experience is emerging as a promising field~\citep{qiao2019WebARPromising}. The World Wide Web Consortium (W3C) has released WebXR, a web-based API that enables the development and experience of XR applications in web browsers\footnote{\url{https://www.w3.org/TR/webxr/}, \url{https://immersiveweb.dev/}\label{footnote:webxr}}.
%
A typical XR app consists of interconnected \textit{scenes}, analogous to \textit{activities} in Android apps, each representing a unique virtual environment. Using Unity as an example, scenes are composed of \textit{GameObjects} and \textit{components}. GameObjects are the graphic elements that users can interact with, while components provide functionalities to GameObjects (e.g., animation, video playback)~\citep{unity_gameobject}.
%
% The specific runtime behaviours of these objects are defined in source code files, such as \emph{scripts} in Unity, which are attached to corresponding GameObjects to implement custom logic and interactivity.
%
% Despite being built with various frameworks and languages (e.g., C\# for Unity and C++ for Unreal Engine), XR applications are generally composed of a set of interconnected \emph{scenes}, similar to \emph{activities} in Android. Each scene may represent a distinct environment where users can experience and interact with virtual content~\citep{nusrat2021HowDevelopersOptimize}.
%
The hierarchical structure of XR scenes, including object relationships and properties (e.g., behaviours, appearances), is managed using specialised data structures called \emph{scene graphs}~\citep{Walsh2002SceneGraphs}. Although XR apps can be built with different frameworks and languages, this core structure is consistent across platforms.

\begin{figure}
    \centering
    \begin{subfigure}[b]{.5\linewidth}
        \centering
        \includegraphics[width=1\linewidth]{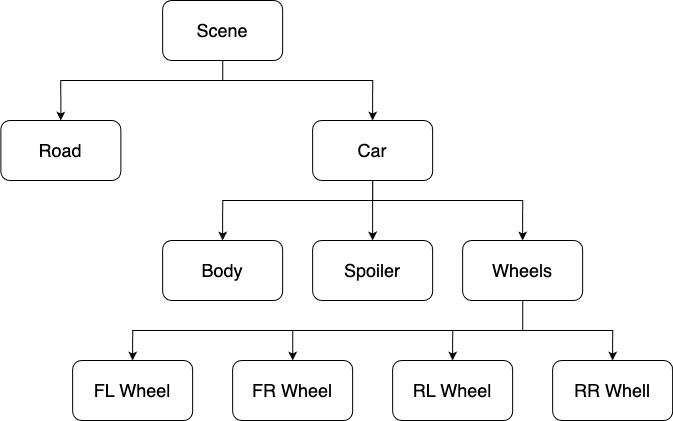}
        \caption{Unity Scene Structure}
        \label{fig:scene_graph_abstract}
    \end{subfigure}
    \hfill
    \begin{subfigure}[b]{.45\linewidth}
        \centering
        \includegraphics[width=1\linewidth]{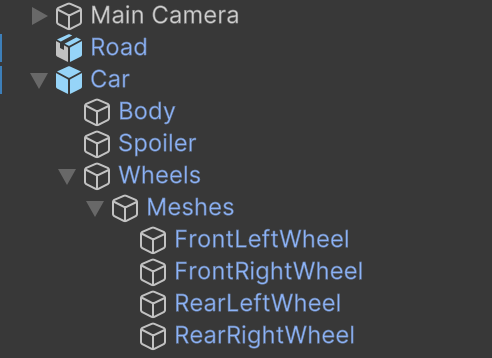}
        \caption{Scene Graph}
        \label{fig:scene_graph_unity}
    \end{subfigure}
    \caption{Examples of Scene Graphs}
    \label{fig:scene_graph}
\end{figure}

% Figure~\ref{fig:scene_graph} illustrates examples of scene graphs. 
Figure~\ref{fig:scene_graph_abstract} depicts the scene graph of an XR environment, including objects such as a \textit{Road} and a \textit{Car}. The \textit{Car} object is further divided into sub-objects like \textit{Body}, \textit{Spoiler}, and \textit{Wheels}. Figure~\ref{fig:scene_graph_unity} presents the corresponding XR scene in the Unity Editor, showcasing the hierarchical relationships among these objects.

% Users of XR apps typically begin at certain positions within the scene, with a camera attached to them, allowing them to view parts of the scene based on their current positions and perspective~\citep{wang2023VRGuideEfficientTesting-PS32}.
%
% They can navigate the scene and adjust their viewpoints to explore its content further.

% \ds{If some of the XR-specific terms, such as scenes, objects, and components, are important, we could explain them using Figure 2.}

\subsubsection{Interaction with XR Applications} \label{sec:xr_interaction}

% comparing XR software and traditional 2D software
In XR apps, user interaction typically involves three main tasks~\citep{kim2020SystematicReviewVirtual, VR/AR}: 
\begin{enumerate*}[label=(\arabic*)]
    \item \textit{navigation:} controlling the user's position and viewing direction within the virtual environment;
    \item \textit{selection:} choosing a point, area, volume, or specific virtual object;
    \item \textit{manipulation:} modifying the parameters of virtual objects, such as their location, orientation, or size.
\end{enumerate*}
Although these tasks are conceptually similar to those in traditional 2D graphical user interfaces, their execution in XR is significantly more complex~\citep{Emery2001CriticalInteractionScenarios, VR/AR}.

This increased complexity arises primarily from the degrees of freedom (DOF) involved—the number of ways an object can move in space. 
In 2D interfaces, interactions typically involve rigid bodies with three DOF: two translations (horizontal and vertical) and one rotation.
In contrast, 3D objects in XR apps operate with six DOF (\emph{6DOF}), encompassing three translational movements (forward/backward, up/down, left/right) and three rotational movements (yaw, pitch and roll).
The additional DOF in XR interactions introduces multiple layers of complexity, demanding more sophisticated interaction techniques and testing methodologies.

% In addition, there are two major functional requirements in XR systems: \textit{occlusion} and \textit{collision}~\citep{Breen2000InteractiveOcclusionCollision, VR/AR}. 
% %
% Occlusion occurs when objects (virtual or physical) block each other from view. It affects how users perceive depth and spatial relationships within the virtual environment, playing a key role in rendering realistic and immersive scenes.
% %
% Collision, on the other hand, refers to the interaction between objects (virtual or physical) when they come into contact. It ensures that virtual objects behave realistically, such as preventing them from passing through one another or accurately responding when interacting with physical objects ~\citep{scheibmeir2019QualityModelTesting-PS20}. 
% % \ds{Is this defined among virtual objects only? What if the collision between physical and virtual objects?}
% %
% Both occlusion and collision are crucial for creating a believable and immersive XR experience, posing challenges that are less significant in traditional 2D interfaces.

% XR interaction devices
The choice of input device is a critical factor in enabling effective interaction within XR environments. Unlike traditional 2D apps, which rely on menus, buttons, and toolbars, XR apps often require specialised hardware to support their unique interaction paradigms.
While mobile and web-based XR apps typically run on conventional devices like smartphones and web browsers, delivering more immersive XR experiences usually demands dedicated devices, such as HMDs, These devices are specifically designed to handle complex and dynamic interactions in XR apps, including 6DOF tracking. This capability enables precise mapping of the user's physical actions, such as movement, rotation, and gestures, into immersive environments, allowing more natural and intuitive interactions.

% These devices are designed to manage the complex and dynamic interactions inherent to XR apps. Most HMDs are equipped with hand-held controllers that support six degrees of freedom (6DOF) tracking\ds{better to explain briefly the six DOFs: roll, pitch, yaw, in addition to movements in x/y/z}. This functionality enables precise mapping of the user's physical actions, such as movement, rotation, and gestures, within the immersive environments.

% interaction sequence, related to testing
In a virtual environment, user inputs are processed in real-time, where even slight variations can significantly alter scene behaviour and the input sequence required to complete a task~\citep{deandrade2023ExploitingDeepReinforcement-PS17}.
This dynamic nature makes reproducing exact input sequences for task replication particularly challenging.
In contrast, traditional GUI software, such as 2D mobile applications, can often be modelled as finite state machines (FSMs)~\citep{su2017GuidedStochasticModelbased}, where input events required to reach a specific state are finite and reproducible. This allows for controlled and predictable interactions.

However, XR apps rely on 6DOF interactions and real-world context, introducing unpredictability. For instance, an XR app may present varying virtual content depending on the user's current location or physical surroundings, making it far more complex to test and replicate specific input sequences compared to traditional 2D GUI apps.

Similarly, although XR apps and 3D video games share common foundations, such as development with the same 3D engines~\citep{bouvier2008CrossBenefitsVirtual}, they differ significantly in interaction mechanisms and real-world integration. This systematic mapping study distinguishes XR apps from 3D video games, acknowledging their shared technological roots but unique user experiences and testing challenges.

\subsection{Software Testing}

Software testing is a practical engineering activity in software development, aimed at ensuring the quality of a software system by evaluating the system under test (SUT)~\citep{Ammann_Offutt_2008}. 

% test case
A central element of software testing is the \textit{test case}, which specifies the conditions for executing the SUT in a certain way. Test cases typically include inputs, execution conditions, and expected results, known as \textit{oracles}, to validate the software's behaviour~\citep{swebok2024, barr2015OracleProblemSoftwarea} and detect faults. 

Testing spans multiple levels, each with distinct objectives:
\begin{enumerate*}[label=(\arabic*)]
    \item \textit{unit testing} verifies individual components, such as methods or classes, in isolation;
    \item \textit{integration testing} examines interactions between components, such as method calls across modules;
    \item \textit{system testing} assesses overall behaviour, including non-functional requirements like security and usability.
\end{enumerate*}
These levels ensure comprehensive evaluation, targeting specific aspects of the system's design and functionality.

Manual testing is time-consuming and resource-intensive, making it impractical for exhaustive testing in large programs. While human testers are indispensable for tasks requiring creativity or domain knowledge, automated testing is increasingly relied upon to streamline repetitive tasks and enhance test coverage.

\subsubsection{Automated Testing and Test Automation}

% Automated testing reduces the reliance on manual effort by automating the creation and execution of test cases. This approach improves efficiency, consistency and thoroughness in testing, particularly for complex and large-scale systems.

\emph{Automated testing} and \emph{test automation} are related terms in the testing domain. We acknowledge these terms might have diverse definitions across academia and industry. For clarity and consistency in this paper, we define automated testing as the automation of both test \emph{generation} and \emph{execution}, while test automation refers solely to the automation of test execution (e.g., driven by manually created test data).

Automated testing reduces the reliance on manual effort by automating the creation and execution of test cases. This approach improves efficiency, consistency and thoroughness in testing, particularly for complex and large-scale systems. Test oracles are an essential part of automated testing and generating accurate and robust oracles is a challenging problem~\citep{molinaTestOracleAutomation2025}.

On the other hand, test automation often relies on manually crafted test data, involving script-based testing frameworks. For instance, tools like \textsc{Espresso} and \textsc{UIAutomator} are scripted-based testing frameworks for Android apps, offering intuitive GUI testing approaches for developers~\citep{gu2023ScriptedGUITesting}.

% Complex domains like GUI apps can significantly benefit from automated testing, where automated tools can explore diverse scenarios, simulate real-world interactions, and validate expected outcomes systematically.

Complex domains like GUI apps can significantly benefit from automated testing and test automation, as these approaches can systematically simulate real-world interactions and validate expected outcomes.

% By addressing the challenges of scalability and repeatability, automated testing has become essential in modern software development, complementing manual efforts to ensure comprehensive quality assurance.

By addressing the challenges of scalability and repeatability, automated testing and test automation have become essential in modern software development, complementing manual efforts to ensure comprehensive quality assurance.

\subsubsection{GUI Testing} \label{sec:background:GUI_testing}

System testing is crucial for GUI apps, complementing unit testing by focusing on user interactions to ensure the software meets requirements and quality standards.

% 2D UI testing
In 2D GUI apps such as Android apps, system testing typically treats the apps as a black box, interacting with the GUI widgets to validate functionality~\citep{kong2019AutomatedTestingAndroid}. Test automation for these apps can be classified into three generations based on the abstraction level of GUI elements~\citep{ardito2019EspressoVsEyeAutomate}:
\begin{enumerate*}[label=(\arabic*)]
    \item \textit{coordinate-based}: interactions rely on exact screen coordinates;
    \item \textit{layout-based}: GUI elements are identified by properties like unique IDs;
    \item \textit{image recognition-based}: components are identified through image matching.
\end{enumerate*}
On the other hand, test generation tools are developed to automatically generate the test cases and execute them within the SUT. They typically create interaction sequences like button clicks or text input to systematically explore the SUT and validate functionality against expected behaviours.

% 3D UI testing
Testing 3D GUIs, such as those found in video games and XR apps, presents significantly greater challenges due to their fine-grained interactivity and complex spatial relationships, which renders automated testing particularly difficult~\citep{politowski2022AutomatedVideoGame}
Emerging methods leveraging advanced techniques--such as evolutionary algorithms and reinforcement learning--aim to systematically explore the 6DOF spaces. However, substantial challenges remain, such as handling flexible camera movements to adjust the user's point of view~\citep{zheng2019WujiAutomaticOnline}.

\section{Related Work and Motivation} \label{sec:related_work}

To explore related work on XR software testing, we conducted a preliminary literature search. During the process, we identified a few secondary studies, such as systematic mapping studies and literature reviews, vaguely related to testing XR applications.
For example, \citet{borsting2022SoftwareEngineeringAugmented} conducted an informal review of software engineering techniques for AR apps, analysing their applicability across various engineering phases, including requirement engineering, implementation and testing.
With regard to testing, the study emphasised the importance of interaction testing and test automation for AR user interfaces. However, it noted significant challenges, such as the lack of formal definitions for AR-specific interactions and the need for tailored testing approaches for unique AR components (e.g., animations, transformations). These challenges pose significant obstacles to achieving effective automated interaction testing.
Additionally, the study discussed the prevalent reliance on user-based usability testing in AR, which often involves labour-intensive user studies. This reliance underscores the need for more reproducible and efficient testing approaches to reduce manual effort and improve scalability.
While this study offered a broad overview of software engineering for AR, our work specifically focuses on the unique challenges and methodologies of software testing for XR apps (i.e., including but not limited to AR).

\citet{kuriUnderstandingSoftwareQuality2021} conducted a mapping study focusing on software quality metrics for validating VR products (e.g., code quality, audio quality, quality of experience) rather than software testing techniques.
The study found that the existing metrics are primarily tailored to specific app types, such as educational apps, making them less applicable to other domains like manufacturing for instance.
Researchers tend to develop custom quality metrics and evaluation methodologies, highlighting the need for a general framework that assesses VR app quality across various dimensions (e.g., code, video, audio).
With the majority of existing metrics focused on the quality of experience and relying on manual evaluation, the study highlights the need for automated, objective methods or metrics to assess software quality.

\subsection{Usability of XR Applications} \label{sec:related:usability}

% usability
While our preliminary literature search did not uncover any systematic review dedicated explicitly to XR software testing, we did find several studies evaluating XR system's \emph{usability}, i.e., the ease of use of specific software systems~\citep{Hertzum2020UsabilityTesting}.
% and \emph{user experience} (UX) of XR systems. These concepts overlap but emphasise different aspects of user interaction. \ds{Why do we emphasise the difference here?}
% Usability focuses on the \emph{effectiveness}, \emph{efficiency} and \emph{satisfaction} within a specified context of use. User experience encompasses a broader scope, emphasizing \emph{subjective perception} and \emph{responses}, such emotions, physical and psychological reactions~\citep{Hertzum2020UsabilityTesting}.
%
% In summary, usability highlights the ease of use, while UX emphasises the user’s overall perception and interaction with the product.

\citet{ramaserichandra2019ReviewUsabilityPerformance} reviewed usability and performance evaluation in VR systems. They identified key usability issues including \emph{health and safety issues}, \emph{social issues} and \emph{sensory constraints}. The study also identified usability evaluation methods, such as cognitive evaluation~\citep{Brown-Johnson2015congnitiveEvaluation}, user analysis~\citep{BARBIERI2017101}, and group testing~\citep{Chen2013groupTesting}.

\citet{dey2018SystematicReview10} conducted a systematic review of AR usability studies from 2005 to 2014. They analysed 369 user studies across various application domains such as education, entertainment, and industry. The most common data collection method was questionnaires, resulting in subjective ratings being the most widely used measure. 
\citet{kim2020SystematicReviewVirtual} reviewed VR systems from a human-computer interaction (HCI) perspective. The findings aligned with those of \citet{dey2018SystematicReview10}, highlighting \emph{subjective measures} as the dominant approach for evaluating VR/AR usability.

Both \citet{ramaserichandra2019ReviewUsabilityPerformance} and \citet{kim2020SystematicReviewVirtual} identified \emph{cybersickness} as a significant usability issue in XR systems. Cybersickness, a form of visually-induced motion sickness experienced in immersive environments, manifests through symptoms such as nausea, disorientation and headaches~\citep{davis2014SystematicReviewCybersickness}.
Various factors may cause cybersickness regarding individuals (e.g., illness, posture), devices (e.g., lag, calibration), and tasks (e.g., control, duration)~\citep{davis2014SystematicReviewCybersickness}.
Studies aiming to comprehensively assess cybersickness by employing subjective and/or objective measures exist.
Subjective measures, e.g., the Simulator Sickness Questionnaire (SSQ)~\citep{Robert1993SSQ}, evaluate participants' self-reported symptoms. Objective measures, in contrast, primarily involve real-time physiological data collection such as heart rate variability (HRV) or eye tracking while participants perform specific tasks~\citep{kaminska2022UsabilityTestingVirtual, qu2022BiophysiologicalsignalsbasedVRCybersickness, kundu2023VRLENSSuperLearningbased}.

\citet{yangMachineLearningMethods2022} conducted a systematic review focusing on the use of machine learning (ML) techniques to study cybersickness. The review examined 26 studies that utilised ML approaches with biometric and neuro-physiological signals, such as electroencephalogram (EEG) and electrocardiogram (ECG) data obtained from wearable devices, for the automated detection of cybersickness.

These studies emphasise the unique usability challenges posed by XR systems compared to traditional software. Testing user interactions in XR is essential, especially because human behaviour in these environments is highly complex and cannot be mathematically modelled to guarantee predictable outcomes~\citep{VR/AR}.
While automated approaches show potential in addressing usability issues like cybersickness detection, they largely depend on user involvement. This reliance on manual testing or \emph{live} user data collection is time-consuming and costly.

Although these findings highlight the importance of \emph{user-centric} evaluations, they also expose a gap in exploring \emph{software-centric} testing approaches. Unlike user-centric methods, software-centric testing can detect failures earlier in the development process and offer more efficient, systematic, and automated testing capabilities. Our mapping study seeks to bridge this gap by examining studies that address usability issues from a software-centric perspective.

\subsection{Motivation}

There is a noticeable gap in the literature regarding a comprehensive overview of testing practices for XR applications.
While recent research on VR app testing highlights the scarcity of literature on software engineering practices specifically for the VR domain~\citep{deandrade2023ExploitingDeepReinforcement-PS17}, this observation extends to the broader XR domain, which remains significantly underexplored.
The gap underscores the need for a formal and in-depth mapping study to analyse existing evidence on XR app testing challenges and techniques, identify research gaps, and suggest further research directions, potentially including systematic literature reviews on specific aspects of XR testing.

Most secondary studies related to XR software testing have primarily focused on usability, a trend that aligns with the findings of \citet{kuriUnderstandingSoftwareQuality2021}. While usability is important for XR experiences, this limited focus has a significant gap in understanding the broader landscape of XR software testing methodologies.
To fill this gap, we conduct this systematic mapping study to provide a comprehensive overview of software testing methodologies for XR apps. Our focus is on techniques and frameworks that prioritise \emph{software requirements}, specifically addressing XR software testing challenges.

Our study adopts an inclusive approach by incorporating empirical studies that, while not directly proposing testing methods, offer valuable insights or theories beneficial to testing. For example, studies analysing common bug types in XR apps provide foundational knowledge that can inform the development of testing strategies. This broader inclusion ensures a more holistic understanding of XR software testing.

% We also observed during our literature search that many studies tend to focus on a particular XR technology, such as VR or AR, without explaining why the proposed techniques are limited in one and not applicable to the other.
% %
% In this study, we aim to map testing techniques relevant to all types of XR applications, including both VR and AR, to explore whether techniques applied in one technology can be extended to others, without focusing on direct comparisons between them.

\section{Mapping Study} \label{sec:methodology}

\begin{figure}
    \centering
    \includegraphics[width=0.9\linewidth]{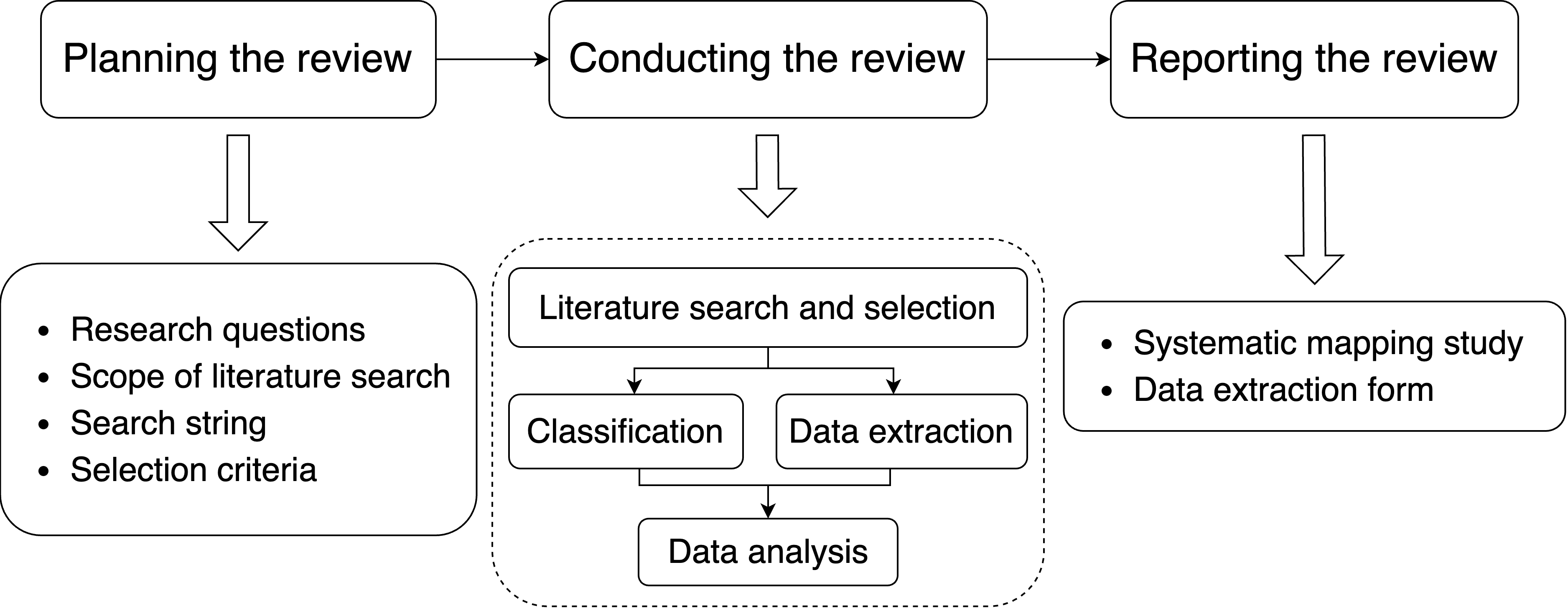}
    \caption{Process of the mapping study}
    \label{fig:mapping_process}
\end{figure}

This systematic mapping study follows the guidelines proposed by \citet{petersen2015GuidelinesConductingSystematic} and is inspired by other systematic mapping studies~\citep{zein2016SystematicMappingStudy, zhang2023FindingCriticalScenarios}. As shown in Figure~\ref{fig:mapping_process}, our mapping process consists of three stages:
\begin{enumerate*}[label=(\arabic*)]
    \item planning, where the research questions and the scope of the literature search are formulated,
    \item conducting, where the authors specify a search strategy, search, and select primary studies, then apply classification and data extraction processes to them subsequently,
    \item reporting the mapping, presenting the outcomes of the study, with complete details of primary studies and extracted data available in the appendix and repository.
\end{enumerate*}

\subsection{Planning the Mapping}

\subsubsection{Research Questions} \label{sec:rqs}

This study aims to develop a comprehensive classification scheme by analysing relevant evidence and insights from the existing literature on software testing for XR applications. 
The scope extends beyond the studies that introduce novel testing techniques for XR apps, encompassing a broader range of research, including empirical studies that provide valuable information for XR software testing (e.g., analysing common bug types within XR software).
Moreover, the study seeks to identify research gaps and challenges and outline future research directions. 
We therefore formulate the following research questions (RQs):

\begin{description}[leftmargin=*]%[style=multiline,leftmargin=3cm]
    \item[\textbf{RQ1:}] \textit{\textbf{What is the current status of XR application testing research?}}
    This question provides an overview of the current landscape in XR software testing research. 
    It will explore general aspects such as the number of publications over recent years, major publication venues, and common research types.
    Additionally, we will investigate the most discussed and emerging topics within XR testing research and which XR technologies (e.g., VR, AR) are the primary focus of current research.
    \item[\textbf{RQ2:}] \textit{\textbf{What are the test facets involved in XR applications?}}
    This research question aims to provide a comprehensive overview of software testing practices for XR applications, including test activities, concerns, and techniques.
    \begin{description}[itemindent=0.5em,leftmargin=1em]
        \item[\textbf{RQ2.1:}] \textit{\textbf{What test activities are involved in XR applications?}}
        This sub-question seeks to identify and categorise the test activities relevant to XR app testing, such as test data generation and test execution.
        By exploring these activities, we aim to understand the current practices in XR app testing and identify potential areas for improvement.
        \item[\textbf{RQ2.2:}] \textit{\textbf{What are the primary test concerns in XR applications?}}
        This sub-question focuses on the key concerns in testing XR apps, particularly the \emph{objectives} (e.g., verifying functionality, improving usability) and \emph{targets} (e.g., user interfaces, XR-specific requirements) of testing. 
        Understanding these concerns helps to clarify the goals and challenges in XR app testing.
        \item[\textbf{RQ2.3:}] \textit{\textbf{What test techniques are employed in XR applications?}}
        This sub-question investigates the specific testing techniques used on XR apps, such as random testing, mutation testing, and model-based testing.
        By analysing these techniques, we aim to investigate the common testing approaches for XR systems.
        \end{description}
    \item[\textbf{RQ3:}] \textit{\textbf{To what extent are XR testing approaches validated?}}
    This RQ explores how the testing methodologies for XR apps are validated. 
    We assess the metrics and environments (e.g., simulation or real devices) used to evaluate their effectiveness.
    
\end{description}

\subsubsection{Search String} \label{sec:search_string}

\begin{table}
    \rowcolors{1}{white}{gray!20}
    \renewcommand{\arraystretch}{1.2}
    \centering
    \caption{PICOC criteria applied to this study}
    \begin{tabular}{r|l} \toprule
         \textbf{Criterion}& \textbf{Description}\\ \midrule
         Population& XR-related software\\ 
         Intervention& Testing techniques or relevant studies addressing testing aspects\\ 
         Comparison& Not applicable\\ 
         Outcome& Insights into methodologies or practices for testing XR applications\\ 
         Context& Peer-reviewed publications\\ \bottomrule
    \end{tabular}
    \label{tab:PICOC}
\end{table}

As the research questions aim to investigate the current research status of XR software testing, it is possible that some studies do not directly focus on testing techniques but instead analyse other aspects related to testing. For example, some studies may investigate the characteristics or challenges of XR systems, such as identifying issues or limitations in XR apps, which can indirectly inform testing practices. 
Specifically, we tackle this by also including studies that explore the nature of \emph{bugs} (or \emph{faults}, etc.) in XR apps, aiming to collect studies analysing them or proposing techniques to detect them. Since XR software testing is still in its early stages, identifying and understanding such issues may still be underexplored.

To ensure that the search process identifies primary studies addressing the RQs, we followed the guidelines by~\citet{kitchenham2007guidelines} to break down the research questions into individual facets using the PICOC model (population, intervention, comparison, outcomes, and context), which then serves as the foundation for designing the search query. The PICOC model is defined in Table~\ref{tab:PICOC} and based on this model, the search query for the digital libraries is:
\begin{center}
    \textit{Search String} = ($\$XR$ \textbf{AND} $\$XR_{acr}$) \textbf{AND} ($\$T$ \textbf{OR} $\$B$)
\end{center}
Here, $\mathit{\$XR}$ denotes the synonyms of \emph{extended reality}; $\mathit{\$XR_{acr}}$ are the acronyms corresponding to these terms; $\mathit{\$T}$ represents the synonyms of \emph{testing}; and $\mathit{\$B}$ are the synonyms of \emph{bugs}. The synonyms used in the search query are detailed in Table~\ref{tab:synonyms}.

\begin{table}
    \rowcolors{1}{white}{gray!20}
    \renewcommand{\arraystretch}{1.2}
    \centering
    \caption{Synonyms in $\mathit{\$XR}$, $\mathit{\$XR_{acr}}$, $\mathit{\$T}$, and $\mathit{\$B}$}
    \begin{tabularx}{\textwidth}{cXc} \toprule
         & \textbf{Synonyms} & \textbf{Metadata} \\ \midrule
         $\mathit{\$XR}$& ``virtual reality'' \textbf{OR} ``augmented reality'' \textbf{OR} ``mixed reality'' \textbf{OR} ``extended reality''& title, full text \\
         $\mathit{\$XR_{acr}}$& VR \textbf{OR} AR \textbf{OR} XR \textbf{OR} MR& title \\ 
         $\mathit{\$T}$& test \textbf{OR} detect \textbf{OR} detection \textbf{OR} verify \textbf{OR} verification& title \\
         $\mathit{\$B}$ & bug \textbf{OR} fault \textbf{OR} defect \textbf{OR} error& title \\ \bottomrule
    \end{tabularx}
    \label{tab:synonyms}
\end{table}

The search string is searched with the studies' titles, and $\mathit{\$XR}$ is additionally searched with full text to ensure $\mathit{\$XR_{acr}}$ in the titles genuinely referred to extended reality. For example, MR also stands for ``magnetic resonance''~\citep{MR-irrelevant-example}, which would yield irrelevant results. By structuring the search string this way, we avoid retrieving extraneous findings related to unrelated fields like medicine.
% \jr{This is not very clear. How does searching for MR in the full text ensure that "MR" (for magnetic resonance) is excluded? I attempted to rewrite the paragraph as: ``The search string is queried within the studies' titles, and $\mathit{\$XR_{acr}}$ is additionally searched within full texts to ensure the acronyms are in topic. For example, MR also stands for ``magnetic resonance'' in medicine, which would yield irrelevant results.'', but then realised the meaning is not the same}

\subsubsection{Search Evaluation} \label{sec:search_evaluation}

To evaluate the quality of the search string, we follow the guidelines of~\citet{petersen2015GuidelinesConductingSystematic}, using a test set of relevant papers, all of which should be found by the search string.
We identified eight studies during our initial literature review
~\citep{wang2022VRTestExtensibleFramework-PS33, 
wang2023VRGuideEfficientTesting-PS32, 
rzig2023VirtualRealityVR-PS31, 
rafi2023PredARTAutomaticOracle-PS23, 
bierbaum2003AutomatedTestingVirtual-PS9, 
correasouza2018AutomatedFunctionalTesting-PS5,
liExploratoryStudyBugs2020-PS6,
andradeUnderstandingVRSoftware2020-PS29}
to compose the test set. 
These studies cover different testing aspects (e.g., functionality, usability, empirical studies), ensuring that the search results include studies with diverse focuses.
We refine the search string iteratively until the search results contain all the studies from the test set.

\subsubsection{Digital Library} \label{sec:digital_library}

To cover as many relevant studies as possible, we conduct our search using OpenAlex\footnote{\url{https://openalex.org/}}, a bibliographic database that indexes scientific papers from major digital libraries, including IEEE Xplore Digital Library\footnote{\url{https://ieeexplore.ieee.org/}}, 
ACM Digital Library\footnote{\url{https://dl.acm.org/}}, and
Scopus\footnote{\url{https://www.scopus.com/}}~\citep{priem2022openalexfullyopenindexscholarly}.
OpenAlex's filter features allow us to restrict the search to studies in the fields of \textbf{Computer Science} and \textbf{Engineering}, reducing irrelevant search results~\citep{zein2016SystematicMappingStudy, tramontana2019AutomatedFunctionalTesting}. 
This restriction excludes papers focused on the applications of XR in other disciplines, such as \textbf{Medicine} and \textbf{Social Sciences}.

\subsubsection{Selection Criteria} \label{sec:selection_criteria}

After executing the search string in the digital library, a list of potentially relevant studies is retrieved. To ensure that we only include studies aligned with the mapping study's objectives and capable of answering the research questions, we developed a set of selection criteria~\citep{petersen2008SystematicMappingStudies}.
As suggested by \citet{petersen2015GuidelinesConductingSystematic}, we piloted the selection criteria (using a sample of 100 studies from the search results) and refined them until consensus was reached among the three authors of this paper that the criteria effectively included relevant studies and excluded irrelevant ones.
As a result of this process, we applied the following inclusion criteria (ICs) and exclusion criteria (ECs):
\begin{description}
    \item[IC1:] Studies must involve software testing techniques, challenges or limitations for extended reality software applications.
    \item[IC2:] Studies published between January 2000 to July 2024.
    \item[IC3:] Studies written in English, published in peer-reviewed journals or conference proceedings, and available in full text.
    \item[IC4:] Studies must be primary studies rather than secondary studies such as systematic literature reviews.
    \item[EC1:] The focus of the studies is not testing but other software development aspects, such as analysis, design or implementation.
    \item[EC2:] Studies do not focus on software-related aspects, such as requirements or integrity, but instead emphasise other areas like user perceptions or hardware configurations.
    \item[EC3:] Studies are duplicated in the search results, including extended versions of existing results.
    \item[EC4:] Studies published in the form of abstract or panel discussion.
\end{description}

\subsection{Conducting the Mapping}

To streamline the methodological information, this subsection provides the fundamental and essential information for conducting this mapping study, including search and selection strategy, classification scheme and data extraction, with more details that can be found at the online repository at \url{https://sites.google.com/view/xr-testing}.

\subsubsection{Search and Selection Strategy} \label{sec:search_selection_strategy}

\begin{figure}
    \centering
    \includegraphics[width=0.8\linewidth]{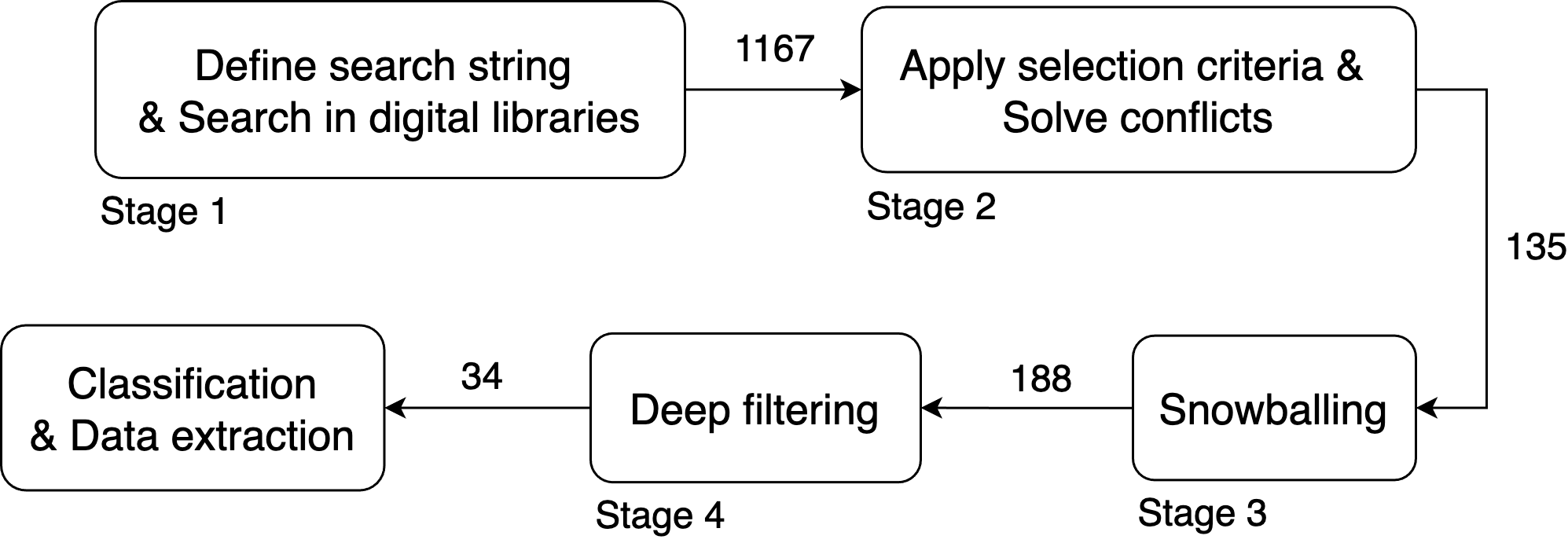}
    \caption{Literature search and selection process. The numbers on the arrows indicate the number of studies provided to the next stage.}
    \label{fig:selection_process}
\end{figure}

The search strategy consists of automated search and manual snowballing. The steps of the search and selection process are illustrated in Figure~\ref{fig:selection_process} and detailed below:

\begin{itemize}
% \item \textbf{\textit{Stage 1:}} The search process begins with the formulation of an initial search string (\S\ref{sec:search_string}).
% The search string is used to retrieve an initial set of potentially relevant studies from the digital library (\S\ref{sec:digital_library}).

\item \textbf{\textit{Stage 1:}} The search process begins with formulating an initial search string (\S\ref{sec:search_string}) to retrieve potential relevant studies from the digital library (\S\ref{sec:digital_library}).

% \item \textbf{\textit{Stage 2:}} To refine the results from the initial search, the selection criteria defined in \S\ref{sec:selection_criteria}\footnote{Excluding IC2 and IC3 which are applied automatically using OpenAlex's filtering feature. The full filters applied by OpenAlex can be found at \url{https://bit.ly/3zTxwCS}.} are applied by reviewing the title and abstract of each study. Based on relevance, each study is labelled as either \emph{``include''}, \emph{``exclude''} or \emph{``unclear''}.
% To minimise bias (as advised by~\citet{petersen2015GuidelinesConductingSystematic}), the three authors independently review each study and the \emph{most inclusive} decision rules are used:
% Only studies unanimously excluded by all authors are discarded immediately; Studies with conflicting labels are considered 
% \emph{``conflicts''} and resolved through a consensus meeting among the authors.
% This approach ensures rigour and reduces the risk of overlooking relevant studies, as further detailed in \S\ref{sec:threats_validity}.

\item \textbf{\textit{Stage 2:}} Selection criteria from \S\ref{sec:selection_criteria}\footnote{Excluding IC2 and IC3 which are applied automatically using OpenAlex's filtering feature. The full filters applied by OpenAlex can be found at \url{https://bit.ly/3zTxwCS}.} are applied by reviewing titles and abstracts, with three authors independently evaluating each study and resolving conflicts through consensus meetings to ensure rigour.

% \item \textbf{\textit{Stage 3:}} Backward snowballing is conducted on the studies retained after applying the selection criteria~(\cite{petersen2015GuidelinesConductingSystematic}). This stage is performed in parallel with deep filtering, classification, and data extraction.
% %
% Following the guidelines by~\citet{wohlin2014GuidelinesSnowballingSystematic}, each study's reference list is reviewed in two steps:
% %
% \begin{enumerate*}[label=(\arabic*)]
%     \item basic exclusion: exclude studies that do not meet basic criteria, such as language, publication year, and venue;
%     \item relevance assessment: evaluate relevance based on titles and context in which each study is referenced, excluding irrelevant works.
% \end{enumerate*}
% %
% Snowballing is an iterative process aimed at minimising the risk of missing relevant studies, as further detailed in \S\ref{sec:threats_validity}; we conducted three iterations of snowballing, ensuring comprehensive coverage of potential primary studies.

\item \textbf{\textit{Stage 3:}} Backward snowballing is conducted on retained studies following guidelines by~\cite{wohlin2014GuidelinesSnowballingSystematic}, with exhaustive iterations performed to ensure comprehensive coverage.

% \item \textbf{\textit{Stage 4:}} Some selection criteria, particularly \emph{E1} and \emph{E2} (\S\ref{sec:selection_criteria}), required a detailed evaluation beyond titles and abstracts. In this stage, the retained studies undergo a full-text review, focusing specifically on exclusion criteria \emph{E1} and \emph{E2} to ensure only relevant studies are given to the following stages.

\item \textbf{\textit{Stage 4:}} Full-text reviews are conducted on remaining studies, focusing specifically on exclusion criteria \emph{E1} and \emph{E2} (\S\ref{sec:selection_criteria}) to ensure only truly relevant studies proceed to the classification phase.

\end{itemize}

After completing this process, the remaining studies form the \emph{primary studies} for this mapping study. These studies proceed to subsequent phases, including classification and data extraction. The full list can be found in Appendix~\ref{appendix:PS}. 

\subsubsection{Classification Scheme} \label{sec:classification_scheme}

The classification scheme organises the primary studies into broad categories to provide a structured overview of the field~\citep{kitchenham2007guidelines}.
Following the guidelines by \citet{petersen2015GuidelinesConductingSystematic}, we applied topic-independent classification, including publication venue and research type, and topic-specific classification.

For topic-specific classification, we utilised the systematic keywording of abstracts method outlined by~\citet{petersen2008SystematicMappingStudies}, extracting keywords from abstracts (consulting introductions and conclusions when needed) to consolidate them into broader categories. To ensure the reliability of keywording, one author classified all studies, while two others independently classified half each, with disagreements resolved through consensus meetings.

\subsubsection{Data Extraction} \label{sec:data_extaction}

To address the RQs outlined in \S\ref{sec:rqs}, we systematically extracted data from each primary study, following guidelines by~\citet{kitchenham2007guidelines}.
We composed a data extraction form (Table~\ref{tab:data_extraction}), including general publication information and research-question-specific data.
Additionally, for reproducibility and practical application, we identified testing-relevant datasets and tools referenced in the primary studies.

For clarity and consistency, the data extraction form was pilot-tested with our initial study set (\S\ref{sec:search_evaluation}). To mitigate bias, one author extracted data from all primary studies, with the other authors reviewing the results, and any disagreements were resolved through consensus meetings.

After the classification and data extraction processes, we analysed and formulated the retrieved data to address the RQs. The analysis results are presented in \S\ref{sec:results}.

\begin{table}[t]
    \rowcolors{1}{white}{gray!20}
    \renewcommand{\arraystretch}{1.3}
    \centering
    \caption{Data Extraction Form}
    \begin{tabularx}{\textwidth}{p{2.3cm}Xc}
        \toprule
        \textbf{Data Item} & \textbf{Description (and possible values)} & \textbf{RQ} \\ \midrule
        \textbf{Title} & Title of the study & - \\ 
        \textbf{Authors} & Names of the study's authors & - \\ 
        \textbf{Year} & Publication year of the study & \emph{RQ1} \\ 
        \textbf{Venue} & Name of the publication venue & - \\ 
        \textbf{Venue Type} & Type of the venue (e.g., conference, workshop, journal) & \emph{RQ1} \\
        \textbf{Topic} & Primary focus area of the study (e.g., usability testing, automated testing) & \emph{RQ1} \\
        \textbf{Research Type} & Research type of the study (e.g., solution proposal, validation research) & \emph{RQ1} \\ \midrule
        \textbf{Technology} & Immersive technology specified in the study (e.g., XR, VR, AR)& \emph{RQ1} \\
        \textbf{Test Activity} & Specific Test activity involved (e.g., test data generation, test tool development, test execution) & \emph{RQ2.1} \\
        \textbf{Test Objective} & Primary objective of the testing approach (e.g., functionality, usability, security)& \emph{RQ2.2}\\
        \textbf{Test Target} & Focus of the testing approaches (e.g., general, GUI, XR-specific requirements)& \emph{RQ2.2}\\
        \textbf{Test Level} & Scope of the testing activities (i.e., unit testing, integration testing, system testing)& \emph{RQ2.3}\\
        \textbf{Test Type}& Type of testing performed (e.g., black box, white box)& \emph{RQ2.3} \\ 
        \textbf{Test Technique} & Core methodologies used for testing (e.g., search-based testing, mutation testing)& \emph{RQ2.3} \\
        \textbf{Evaluation Environment} & Environment for evaluating the testing approaches (e.g., Unity Editor, HMD, mobile device)& \emph{RQ3} \\
        \textbf{Metrics} & Metrics used to evaluate testing techniques (e.g., coverage, mutation score)& \emph{RQ3} \\ \midrule
        \textbf{Dataset\textsubscript{\textit{train}}}& Details of datasets used for training machine learning-based approaches, including content types (e.g., video, image) and dataset size & \emph{discuss.} \\
        \textbf{Dataset\textsubscript{\textit{eval}}} & Details of datasets for evaluating testing techniques, including content types and size & \emph{discuss.} \\ 
        \textbf{Tool} & Details of software tools proposed or used by the study & \emph{discuss.} \\
        \bottomrule
    \end{tabularx}
    \label{tab:data_extraction}
\end{table}

% \subsubsection{Data Analysis} \label{sec:data_synthesis}

% The purpose of this task is to analyse the classification and extracted data to address the research questions (RQs) outlined in \S\ref{sec:rqs}~\citep{kitchenham2007guidelines}. 
% %
% For instance, to answer RQ1, which investigates the current status of XR software testing research, we analyse the topics and research types derived from the classification, with relevant extracted data to provide a comprehensive overview of the XR software testing research landscape.
% %
% Each RQ is mapped to specific data fields in the data extraction form (see Table~\ref{tab:data_extraction}), ensuring all RQs are adequately addressed. 

% The analysis results and findings are presented in \S\ref{sec:results}. The extracted data are systematically tabulated to align with the RQs, highlighting key similarities and differences across the studies.

\subsection{Reporting the Mapping}

The report contains two parts. The first part comprises this paper, which outlines the study's methodology and findings, and the second part is the data extraction form, which details the raw data collected and the basis for the study's conclusions.
The complete results of the classification and data extraction are publicly accessible at: \url{https://sites.google.com/view/xr-testing}.

\subsection{Threats to Validity} \label{sec:threats_validity}

This section addresses potential threats to the completeness of the literature search and selection process. This threat is influenced by the search string choice, the bibliographic database limitations, and the robustness of the literature selection process.

% literature selection
To mitigate the risk of excluding relevant studies during the selection process, three authors independently screened all search results, following the most inclusive approach. Conflicts were resolved through consensus discussions.
While the inter-rater agreement was not formally measured, our approach prioritised achieving absolute consensus to maximise the inclusion of relevant studies.

% digital libraries
% Although OpenAlex indexes publications from major digital libraries such as IEEE, ACM, and Scopus, this is a risk that relevant studies are missed in OpenAlex. 

We acknowledge potential limitations in our search strategy--solely using OpenAlex as the digital library--may affect the thoroughness of our study. While OpenAlex indexes publications from major digital libraries such as IEEE, ACM, and Scopus, we recognise that it does not index some relevant studies (e.g., those that may exist exclusively in specialised venues or databases). Additionally, very recently published studies might not have been indexed in OpenAlex at the time of our search, creating a temporal bias against the latest research. 

% snowballing
% To mitigate this risk, we adopted an exhaustive iterative strategy. Relevant studies were identified from the reference lists of each included study, and the process was repeated until no new relevant studies were discovered.
%
% In this mapping study, the snowballing process involved three iterations, which ensured that all potentially relevant studies were captured and included.

To mitigate these risks, we adopted an exhaustive iterative snowballing strategy. We systematically identified relevant studies from the reference lists of each included study and repeated this process until no new relevant studies were discovered. Our snowballing process involved three iterations, which significantly reduced the likelihood of missing important contributions to the field.

\section{Results} \label{sec:results}

In this section, we present the results of this mapping study, including information about the search and selection results of the primary studies, and answer the research questions based on the information from the primary studies. 

The initial search returned 1167 studies from OpenAlex. The selection process, as outlined in \S\ref{sec:search_selection_strategy}, reduced this to 135 studies after applying the selection criteria. In parallel with deep filtering, backward snowballing identified 53 additional studies, resulting in a final set of 34 primary studies for this mapping study (see Appendix~\ref{appendix:PS} for the complete list of primary studies).

\begin{figure}
    \centering
    \includegraphics[width=0.9\linewidth]{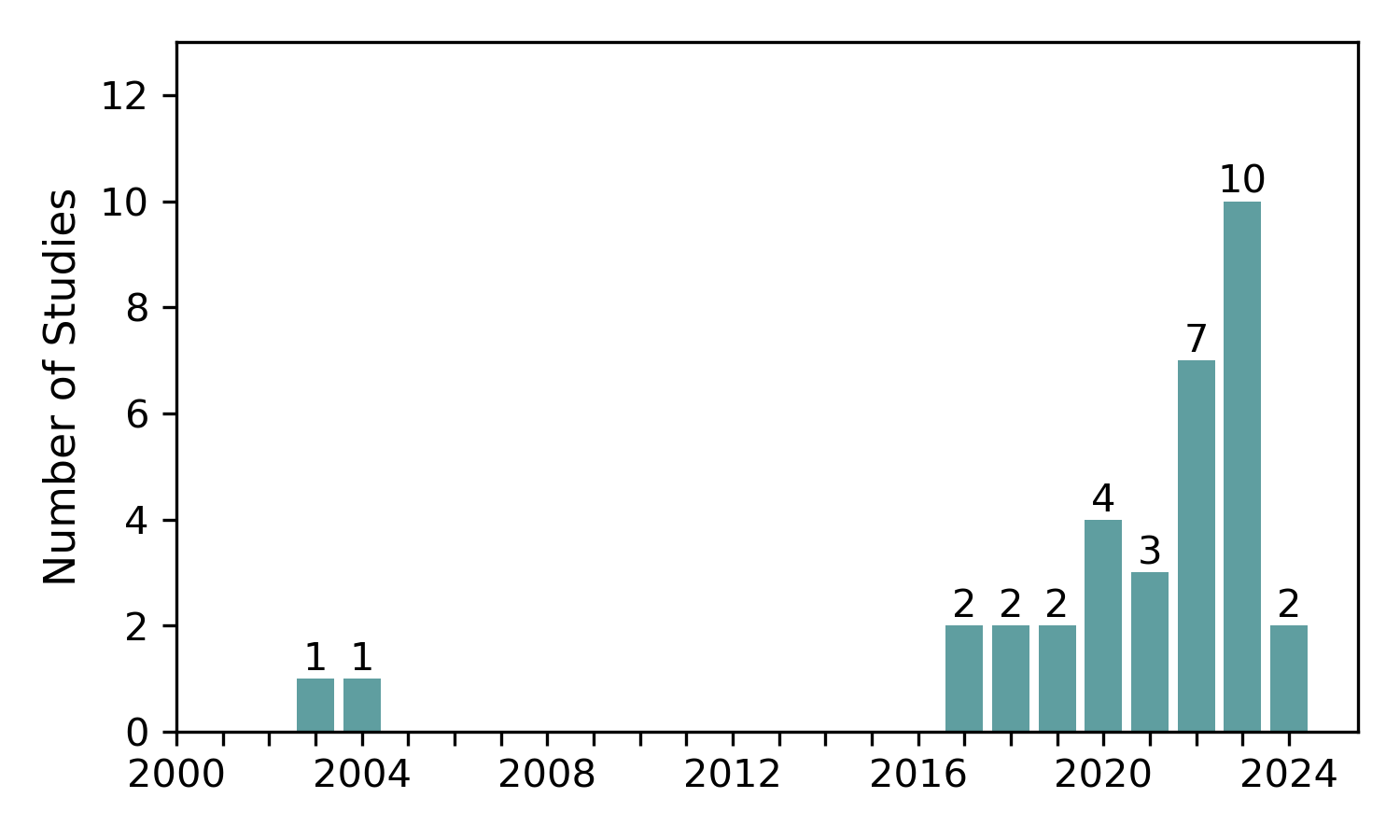}
    \caption{Publication Years of Primary Studies}
    \label{fig:publication_year}
\end{figure}

Figure~\ref{fig:publication_year} shows the publication trend of the primary studies from January 2000 to July 2024. The data reveal that the XR software testing research field was relatively inactive before 2017, with only two studies published.
However, starting in 2017, the number of studies began to increase gradually, reaching a peak in 2023, with ten studies published that year. As the literature search for this study was conducted in July 2024, the number of studies published in 2024 was not completely recorded.

\subsection{RQ1: Research Status}

To address RQ1, which explores the current status of research in XR software testing, we present the classification results of the primary studies. Additionally, we analyse the immersive technologies (e.g., AR, VR) featured as testing subjects in these studies, providing insights into the technologies most frequently explored in this domain.

The studies are classified based on the following criteria:
\begin{enumerate*}[label=(\arabic*)]
    \item venue types (e.g., conferences, journals),
    \item study topics (e.g., automated testing, usability testing), and 
    \item research types (e.g., solution proposal, evaluation research).
\end{enumerate*}

\subsubsection{Venues}

\begin{figure}
    \centering
    \includegraphics[width=0.9\linewidth]{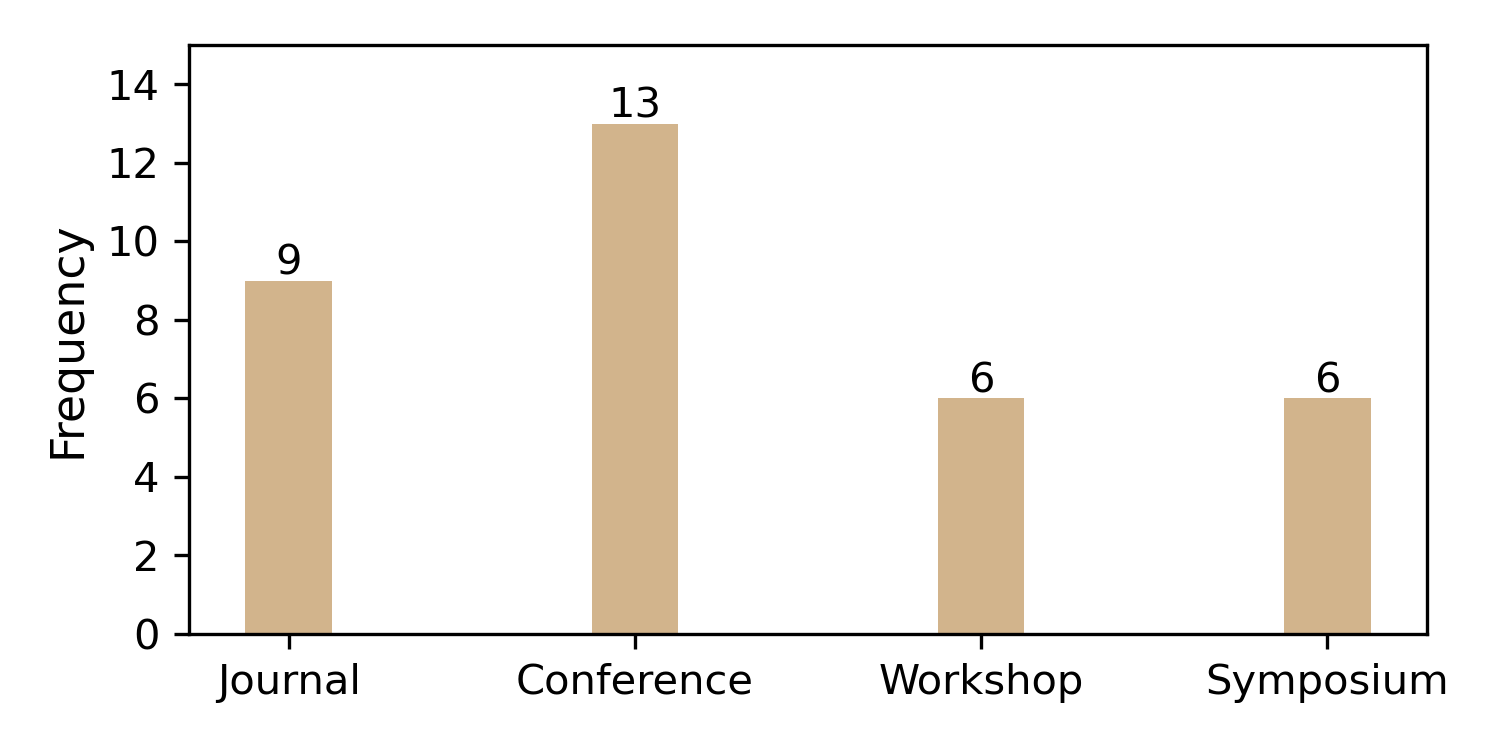}
    \caption{Distribution of publication venue types for the primary studies}
    \label{fig:venue_type}
\end{figure}

The distribution of primary studies by venue type is presented in Figure~\ref{fig:venue_type}. Conferences emerge as the dominant venue type, accounting for about 38\% of the primary studies. Journals are the second most common venue, publishing approximately 26\% of the studies. Workshops and symposiums each represent about 18\% of the total primary studies\footnote{The distinction between symposiums and conferences is not always clear. For classification purposes, venues with ``conference" in their tile are categorised as conferences and those with ``symposium" are classified as symposiums}. 
The complete list of all publication venues for the primary studies is available at: \url{https://sites.google.com/view/xr-testing}.

\subsubsection{Topics} \label{sec:results:topics}

\begin{table}
    \rowcolors{1}{white}{gray!20}
    \renewcommand{\arraystretch}{1.2}
    \centering
    \caption{Research topics in the primary studies}
    \begin{tabular}{llc} \toprule
        \textbf{Topic} & \textbf{Primary studies} & \textbf{Number} \\ \midrule
        XR-specific testing & PS16, PS17, PS20, PS22 & 4 \\
        Scene testing & PS4, PS12, PS32, PS33 & 4 \\
        Security testing & PS2, PS13, PS14, PS15, PS18, PS30 & 6 \\
        Usability testing & PS1, PS7, PS10, PS11, PS19, PS21 & 6 \\
        Automated testing & PS3, PS5, PS8, PS23, PS24, PS25, PS27 & 7 \\
        Test automation & PS9, PS26, PS34 & 3 \\
        Open-source projects & PS6, PS28, PS31 & 3 \\
        Stakeholder survey & PS29 & 1 \\
        \bottomrule
    \end{tabular}
    \label{tab:classification:topic}
\end{table}

As explained in \S\ref{sec:classification_scheme}, we carefully categorise the primary studies into eight topics.
Table~\ref{tab:classification:topic} presents the resulting classification.
Overall, \emph{automated testing} emerges as the most prominent topic, with seven studies (21\% of the total). \emph{Usability testing} and \emph{security testing} each account for six studies (18\%). \emph{XR-specific testing} and \emph{scene testing}, each contribute four studies (12\%), while \emph{test automation} and \emph{open-source projects} each comprise three studies (9\%). Finally, \emph{stakeholder survey} represents the least explored topic, with only one study (3\%).

Below we provide details for each topic, including:
\begin{enumerate*}[label=(\arabic*)]
    \item \textit{description:} a general overview of the topic,
    \item \textit{examples:} summaries of representative studies within the topic, and
    \item \textit{implications:} insights into the potential consequences, proposed solutions, or general guidelines for addressing the challenges associated with the topic.
\end{enumerate*}

\paragraph{XR-specific Testing}
\begin{itemize}
    % \item \textbf{Description:} Testing techniques designed to XR-specific requirements, such as collision, occlusion, registration, and tracking~\citep{VR/AR}. 
    \item \textbf{Description:} Testing techniques for XR-specific requirements like collision, occlusion, registration, and tracking~\citep{VR/AR}.
    % \item \textbf{Examples:} Collision and occlusion are critical real-time requirements in XR systems. Collision refers to the interaction between objects (virtual or physical) when they come into contact and occlusion occurs when objects (virtual or physical) block each other from view~\citep{Breen2000InteractiveOcclusionCollision, VR/AR}. Testing these aspects ensures realistic interactions between virtual and real objects.
    % PS17~\citep{deandrade2023ExploitingDeepReinforcement-PS17} proposes an approach that uses automatic test data generation to explore virtual scenes and detect collision and camera occlusion faults in VR apps.

    \item \textbf{Examples:} Collision and occlusion are critical real-time requirements in XR apps, where the former refers to the interaction between objects when they come into contact and the latter occurs when objects block each other from view~\citep{Breen2000InteractiveOcclusionCollision, VR/AR}. Testing these aspects ensures realistic interactions between virtual and real objects.
    PS17~\citep{deandrade2023ExploitingDeepReinforcement-PS17} proposes an approach that automatically generates test data to detect incorrect collision and occlusion in VR apps.
    
    % \item \textbf{Implications:} Testing XR-specific requirements demands a deep understanding of their impact on software behaviour, which calls for novel testing techniques tailored to these unique challenges.
    
    \item \textbf{Implications:} Testing XR-specific requirements demands a deep understanding of their impact on software behaviour, and tailored testing techniques for these unique challenges.
\end{itemize}

\paragraph{Scene Testing}
\begin{itemize}
    % \item \textbf{Description:} Focuses on validating XR functionality by \emph{exploring XR scenes} through interacting with virtual objects and navigating within virtual scenes.

    \item \textbf{Description:} Validates XR functionality through exploration of XR scenes and interaction with virtual objects.
    
    % \item \textbf{Examples:} PS32~\citep{wang2023VRGuideEfficientTesting-PS32} and PS33~\citep{wang2022VRTestExtensibleFramework-PS33} introduced testing techniques for VR scene testing, focusing on exploring virtual environments, triggering interactable virtual objects, and optimising interaction routes to enhance exploration efficiency.

    \item \textbf{Examples:} PS32~\citep{wang2023VRGuideEfficientTesting-PS32} and PS33~\citep{wang2022VRTestExtensibleFramework-PS33} introduced VR scene testing techniques, focusing on exploring environments, triggering interactable objects, and optimising interaction routes.
    
    % \item \textbf{Implications:} Scene testing extends principles of 2D GUI testing into 3D environments, targeting interactable components. As discussed in \S\ref{sec:xr_interaction}, unlike 2D apps, XR apps rely on 6DOF interactions, adding complexity to their testing.

    \item \textbf{Implications:} Scene testing extends principles of 2D GUI testing into more complex 3D environments with 6DOF interactions.
\end{itemize}

\paragraph{Usability Testing}
\begin{itemize}
    % \item \textbf{Description:}  Focuses on identifying usability issues in XR software, rather than addressing functional defects.

    \item \textbf{Description:} Identifies usability issues (e.g., side effects) in XR software.
    
    % \item \textbf{Examples:} 
    % PS1~\citep{jung2017360degStereoImage-PS1}, PS19~\citet{li2024LessCybersicknessPlease-PS19}, and PS21~\citep{kim2017MeasurementExceptionalMotion-PS21} proposed approaches for detecting \emph{cybersickness}, a prevalent side effect in VR systems. These methods utilise the visual content, such as screenshots from applications, to analyse and identify potential usability concerns.

    \item \textbf{Examples:} 
    PS1~\citep{jung2017360degStereoImage-PS1}, PS19~\citep{li2024LessCybersicknessPlease-PS19}, and PS21~\citep{kim2017MeasurementExceptionalMotion-PS21} proposed approaches for detecting \emph{cybersickness}, a prevalent side effect in VR systems using visual content (e.g., screenshots) analysis.
    
    % \item \textbf{Implications:} While most existing research employs user-centric methods, such as user studies, as discussed in \S\ref{sec:related:usability}, these studies also analyse the root causes of usability issues. This understanding provides a foundation for developing software-centric approaches to systematically identify and mitigate usability problems in XR systems.

    \item \textbf{Implications:} While most existing research employs user-centric methods, such as user studies (\S\ref{sec:related:usability}), understanding the root causes of these usability issues would enable the development of systematic software-centric approaches.
\end{itemize}

\paragraph{Security Testing}
\begin{itemize}
    % \item \textbf{Description:} Focuses on identifying and mitigating \emph{security} and \emph{privacy} issues in XR software. Security testing aims to detect system intrusion and address vulnerability, while privacy testing protects users' sensitive information.

    \item \textbf{Description:} Identifies and mitigates \emph{security} and \emph{privacy} issues in XR software. Security testing detects system intrusion and addresses vulnerability, while privacy testing protects users' sensitive information.
    
    % \item \textbf{Examples:} PS18~\citep{lehman2022HiddenPlainSight-PS18} introduces a framework to address privacy issues in mobile AR apps, specifically focusing on hidden operations that compromise user data. On the other hand, PS15~\citep{valluripally2023DetectionSecurityPrivacy-PS15} targets attacks designed to disrupt user experiences in VR environments.

    \item \textbf{Examples:} PS18~\citep{lehman2022HiddenPlainSight-PS18} introduces a framework to address privacy issues in mobile AR apps, while PS15~\citep{valluripally2023DetectionSecurityPrivacy-PS15} targets attacks that disrupt VR user experiences.
    
    % \item \textbf{Implications:} Security and privacy testing in XR systems should account for XR-specific features, such as interaction with physical environments and differentiation between real and world objects~\citep{casey2021ImmersiveVirtualReality}.

    \item \textbf{Implications:} Security and privacy testing should account for XR-specific features, such as interaction with physical environments and differentiation between real and virtual objects~\citep{casey2021ImmersiveVirtualReality}.
\end{itemize}

\paragraph{Automated Testing}
\begin{itemize}
    % \item \textbf{Description:} Proposes testing techniques that automate both test \emph{generation} and \emph{execution} for functional testing, without specifically targeting XR-specific or non-functional requirements.

    \item \textbf{Description:} Automates both test \emph{generation} and \emph{execution} for functional testing, without specifically targeting XR-specific or non-functional requirements.

    % \item \textbf{Examples:} PS23~\citep{rafi2023PredARTAutomaticOracle-PS23} and PS27~\citep{yang2024AutomaticOraclePrediction-PS27} tackle the oracle problem of object misplacement in AR apps. These studies analyse screenshots depicting various object misplacement scenarios and use crowdsourcing to generate test data. Neural networks are then employed to automatically detect misplacement errors in the screenshots, achieving full automation for both test data generation and execution.

    \item \textbf{Examples:} PS23~\citep{rafi2023PredARTAutomaticOracle-PS23} and PS27~\citep{yang2024AutomaticOraclePrediction-PS27} tackle the oracle problem of object misplacement in AR apps using neural networks to detect errors in screenshots depicting object misplacement scenarios.
    
    % \item \textbf{Implications:} Effective test automation often requires a thorough understanding of the SUT, particularly its requirements or test oracles. This understanding is crucial for systematically generating reliable test data and ensuring accurate testing outcomes.

    \item \textbf{Implications:} Effective test automation often requires a thorough understanding of system requirements or test oracles, which is essential for systematically generating reliable test data.
\end{itemize}

\paragraph{Test Automation}
\begin{itemize}
    % \item \textbf{Description:} Automating test execution while relying on manually created test data, distinguishing it from \emph{automated testing}, which automates both test generation and execution.

    \item \textbf{Description:} Automating test execution but not test generation.
    
    % \item \textbf{Examples:} PS34~\citep{figueira2022YoukaiCrossPlatformFramework-PS34} presents a unit testing framework for Unity-based VR/AR apps. It enables interaction with UI elements through manually created test scripts.

    \item \textbf{Examples:} PS34~\citep{figueira2022YoukaiCrossPlatformFramework-PS34} presents a unit testing framework for Unity-based VR/AR apps using manually created test scripts.
    
    % \item \textbf{Implications:} Test automation often involves script-based testing frameworks. For instance, tools like \textit{Espresso}\footnote{\url{https://developer.android.com/training/testing/espresso}}, a script-based testing tool for Android apps, exemplify this approach by enabling automated test execution driven by predefined scripts.

    \item \textbf{Implications:} Test automation often involves script-based testing frameworks for automated execution driven by predefined tests.
\end{itemize}

\paragraph{Open-source Projects}
\begin{itemize}
    % \item \textbf{Description:} Empirical studies that analyse open-source XR projects to gain insights into current testing practices and challenges.

    \item \textbf{Description:} Empirical studies that analyse open-source XR projects to gain insights into current testing practices and challenges.
    
    % \item \textbf{Examples:} PS6~\citep{liExploratoryStudyBugs2020-PS6} examines bugs in open-source WebXR projects, building a taxonomy based on symptoms and root causes. Similarly, PS31~\citep{rzig2023VirtualRealityVR-PS31} investigates open-source VR projects, revealing a lack of automated tests and insufficient quality in existing tests.

    \item \textbf{Examples:} PS6~\citep{liExploratoryStudyBugs2020-PS6} examines bugs in open-source WebXR projects to explore bug symptoms and root causes, while PS31~\citep{rzig2023VirtualRealityVR-PS31} investigates open-source VR projects and reveals their insufficiency in testing.
    
    % \item \textbf{Implications:} Studying open-source projects provides valuable empirical data for understanding XR testing practices, particularly in this emerging field. These analyses offer practical recommendations for researchers and practitioners to improve strategies for XR software testing.

    \item \textbf{Implications:} Studying open-source projects provides valuable empirical data and practical recommendations for the emerging field of XR software testing.
\end{itemize}

\paragraph{Stakeholder Survey}
\begin{itemize}
    % \item \textbf{Description:} Surveys and interviews conducted with real-world stakeholders, such as users, developers, and researchers of XR software, aim to gather diverse perspectives on practices and challenges in XR software testing.

    \item \textbf{Description:} Surveys and interviews with real-world stakeholders like XR users and developers, gathering perspectives on testing practices and challenges.
    
    % \item \textbf{Examples:} PS29~\citep{andradeUnderstandingVRSoftware2020-PS29} surveys XR stakeholders to understand software testing practices in VR apps, highlighting key concerns and common faults, such as interaction issues and crashes, within these systems.

    \item \textbf{Examples:} PS29~\citep{andradeUnderstandingVRSoftware2020-PS29} surveys XR stakeholders to understand software testing practices, highlighting key concerns and common faults, such as interaction issues and crashes.
    
    % \item \textbf{Implications:} Stakeholder surveys provide empirical insights grounded in real-world experiences, offering perspectives distinct from those derived from analysing open-source projects. These insights contribute to understanding current challenges and can guide the improvement of XR testing. 

    \item \textbf{Implications:} Stakeholder surveys provide real-world insights, complementing open-source project analysis and guiding testing improvements.
\end{itemize}

\subsubsection{Research Types} \label{sec:research_type}

\begin{table}
    \rowcolors{1}{white}{gray!20}
    \renewcommand{\arraystretch}{1.2}
    \centering
    \caption{Research types in the primary studies}
    \begin{tabular}{lc} \toprule
        \textbf{Research type} & \textbf{Number of studies} \\ \midrule
        Validation research & 2 \\
        Evaluation research & 5 \\
        Solution proposal & 6 \\
        Solution proposal \emph{and} validation research & 15 \\
        Solution proposal \emph{and} evaluation research & 4 \\ 
        Philosophical papers & 2 \\
        \bottomrule
    \end{tabular}
    \label{tab:classification:research_type}
\end{table}

For research types, we adopt the classification categories proposed by~\citet{wieringa2006RequirementsEngineeringPaper}: \textit{solution proposal}, \textit{validation research}, \textit{evaluation research}, \textit{philosophical paper}, \textit{opinion paper}, and \textit{experience paper}. These categories have been carefully reviewed and adapted to align with the scope of XR software testing, ensuring relevance to the primary studies. Notably, a study can span multiple categories, such as studies that propose solutions and include initial validation.

Table~\ref{tab:classification:research_type} summarises the results. \textit{Solution proposal and validation research} emerges as the most prevalent research type, encompassing 15 studies (44\% of the total). \textit{Solution proposal} accounts for 6 studies (18\%), \textit{evaluation research} includes 5 studies (15\%), and \textit{solution proposal and evaluation research} covers 4 (12\%). On the other hand, \textit{validation research} and \textit{philosophical papers} each account for 2 studies (6\%). Notably, no \textit{opinion paper} and \textit{experience paper} were identified in the primary studies.

Similar to the approach in \S\ref{sec:results:topics}, we present the \textit{description} and \textit{examples} of each research type based on the primary studies.

\paragraph{Validation Research}
\begin{itemize}
    \item \textbf{Description:} Providing initial validations of solutions or problems, typically involving limited experiments in controlled, simplified settings, such as toy applications (e.g., research prototypes or low-popularity open-source apps) and datasets of minimal complexity.
    \item \textbf{Examples:} PS12~\citep{richardgunawan2023BlackboxTestingVirtual-PS12} introduces a black-box testing approach for a VR musical instrument game, using equivalence partition to design the test cases. Validation was limited to manual assessment of test results without systematic methodologies.
\end{itemize}

\paragraph{Evaluation Research}
\begin{itemize}
    \item \textbf{Description:} Conducts rigorous testing in real-world settings, addressing meaningful research questions. These studies engage real users or practitioners or evaluate with practical applications, such as industrial software or widely used open-source projects, and using datasets derived from real-world scenarios.
    \item \textbf{Examples:} PS31~\citep{rzig2023VirtualRealityVR-PS31} conducted an empirical study on VR automated testing in open-source VR projects, revealing gaps in current practices.
\end{itemize}

\paragraph{Solution Proposal}
\begin{itemize}
    \item \textbf{Description:} Proposes innovative approaches to XR testing challenges, focusing on theoretical benefits with minimal empirical evidence. These studies typically use basic examples and lack experimental validation with real-world applications.
    \item \textbf{Examples:} PS3~\citep{prasetya2021AgentbasedArchitectureAIEnhanced-PS3} presents an autonomous agent-based testing framework for XR systems. The study details the architecture and potential applications but without experimental assessment.
\end{itemize}

\paragraph{Solution Proposal and Validation Research}
\begin{itemize}
    \item \textbf{Description:}  Combines proposing novel solutions with preliminary validation, typically in simplified experimental settings.
    \item \textbf{Examples}: PS23~\citep{rafi2023PredARTAutomaticOracle-PS23} presents a technique for detecting object misplacement issues in AR apps. It is validated using Unity-provided examples rather than real-world apps.
\end{itemize}

\paragraph{Solution Proposal and Evaluation Research}
\begin{itemize}
    \item \textbf{Description:} Proposes novel solutions and rigorously evaluates them in real-world contexts.
    \item \textbf{Examples}: PS19~\citep{li2024LessCybersicknessPlease-PS19} introduces a technique to detect stereoscopic visual inconsistencies in VR apps, validated using screenshots from real-world VR apps available on the Steam store.
\end{itemize}

\paragraph{Philosophical Papers}
\begin{itemize}
    \item \textbf{Description:} Focuses on theoretical perspectives or conceptual frameworks rather than implementing technical solutions. These studies aim to propose new ways of thinking about challenges, without presenting concrete implementations.
    \item \textbf{Examples:} PS13~\citep{kilger2021DetectingPreventingFaked-PS13} outlines general guidelines for detecting and preventing cybersecurity attacks in MR environments, identifying threats and countermeasures but offering no implementations.
\end{itemize}

\hfill

The majority of primary studies, 25 out of 34 (74\%), propose novel solutions for XR software testing problems. Among these, 15 (60\%) include only basic validation, highlighting the emerging nature of the field, with limited research evaluated in real-world scenarios.

The first empirical evaluation study was published in 2019 (PS26). More recently, 2023 saw the introduction of two novel testing solutions evaluated in real-world contexts (PS16 and PS32). This trend suggests increasing potential for applying XR testing techniques in practical, real-world environments in the near future.

% Figure~\ref{fig:topic_vs_research_type} presents a bubble plot illustrating the distribution of studies across research topics and types. The majority of studies in each topic propose novel solutions and validate them. For instance, in XR-specific testing, 6 out of 12 studies fall under the solution proposal + validation research category, while for usability testing, 4 out of 6 studies belong to this category. These findings highlight the positive contributions of the XR software testing community, showcasing high-quality research that introduces innovative techniques with empirical validation.

% \begin{figure}
%     \centering
%     \includegraphics[width=1\linewidth]{figures/bubble_topic_vs_research-type.png}
%     \caption{Bubble plot for research topics against research type}
%     \label{fig:topic_vs_research_type}
% \end{figure}

\subsubsection{Immersive Technology}

\begin{figure}
    \centering
    \includegraphics[width=0.9\linewidth]{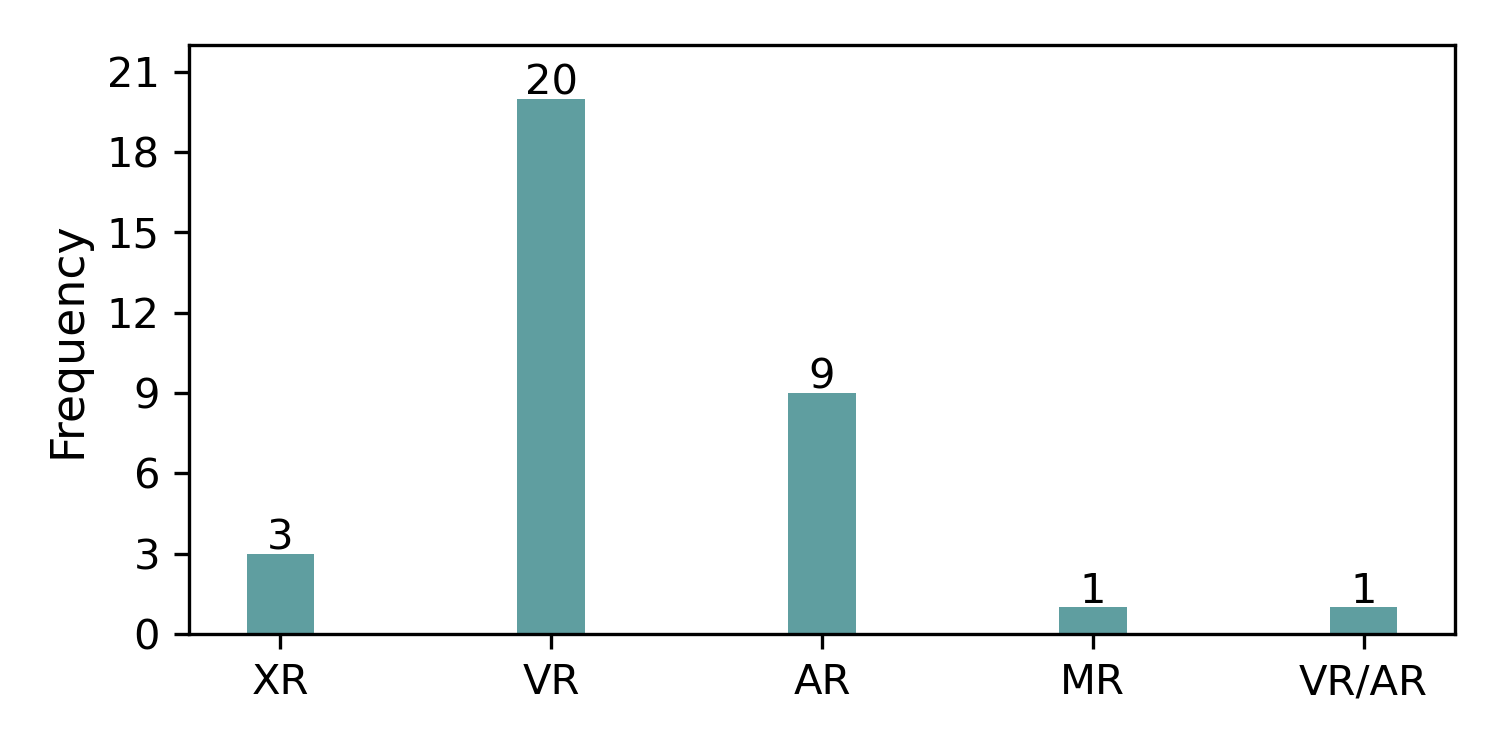}
    \caption{Immersive Technologies of Primary Studies}
    \label{fig:xr_technology}
\end{figure}

We analyse the primary studies by examining the specific immersive technologies targeted for testing in these studies.
Figure~\ref{fig:xr_technology} summarises the technologies examined in the primary studies. About 59\% of the studies focus on testing VR apps, while approximately 26\% target AR apps. Only a small number of studies explore broader or integrated scopes: three focus on XR systems\footnote{We acknowledge XR is an umbrella term that includes VR, AR, and MR; this categorisation is based on each study' specific context and terminology.}, one investigates MR testing, and one addresses both VR and AR testing.

The findings indicate that while VR and AR testing have received significant research attention, studies addressing broader scopes are still in the early stages of development within the research landscape.

\begin{rqanswer}
\textbf{The answer to RQ1, i.e., the current status of XR application testing research, is as follows:}

\textbf{Publication trends:} Research on XR software testing has grown steadily, increasing from 2 publications in 2017 to 10 in 2023.

\textbf{Publication venues:} Conferences and journals are the main publication venues, representing 38\% and 26\% of the studies, respectively.

\textbf{Research topics:} Automated testing is the most prevalent research topic, accounting for 21\% of studies, followed by usability testing and security testing, each contributing 18\%.

\textbf{Research types:} 74\% of studies propose novel XR testing solutions, with 60\% relying on preliminary validations in controlled settings.

\textbf{Immersive technologies:} VR dominates with 59\% of studies and AR represents 26\%. XR, MR, and cross-technology research contributes 15\%.

In summary, XR software testing is an emerging field, steadily gaining momentum.
\end{rqanswer}

\subsection{RQ2: Testing Facets} 

We classify the primary studies based on three key test facets: \emph{test activities} (e.g., test generation, test execution), \emph{test concerns} (including objectives like functionality and security, and targets such as user interaction and collision ), and \emph{test techniques} (e.g., random testing, model-based testing). 
To ensure the meaningfulness of the extracted information, we exclude the studies that do not directly yield testing facets, which are five studies identified as empirical studies (PS2, PS6, PS28, PS29, PS31) and two classified as philosophical papers (PS13 and PS24). The remaining 27 studies are analysed to address this research question.

\subsubsection{RQ2.1: Test Activities} \label{sec:results:test_activity}

\begin{figure}
    \centering
    \includegraphics[width=0.7\linewidth]{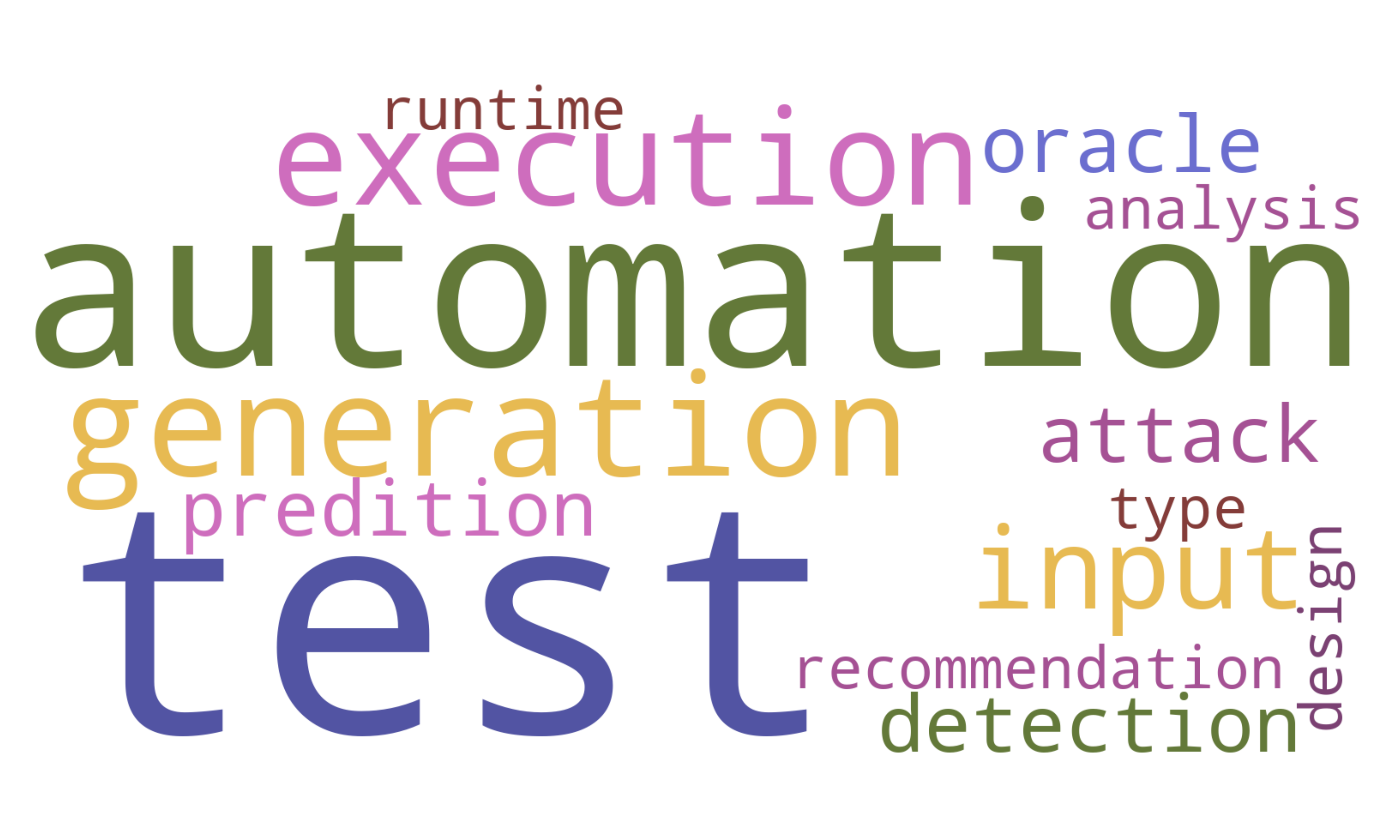}
    \caption{Word cloud based on the test activities of the primary studies}
    \label{fig:test_activities}
\end{figure}

Figure~\ref{fig:test_activities} visualises the distribution of test activities in the primary studies using a word cloud. The size of each keyword in the word cloud corresponds to its frequency, with larger keywords appearing more frequently in the test activities.

For instance, ``test" is the most prominent keyword, reflecting its centrality across various activities. Among these, ``generation" and ``automation" are the next most prominent keywords, indicating that activities such as test generation and test automation are the most frequently addressed activities in the studies.

11 studies include the keywords ``test'' and ``automation'', and all are associated with the test activity \textit{test automation}. On the other hand, eight studies include the keywords ``test" and ``generation", among them, three involve the activity \textit{test generation}, and five involve \textit{test input generation}. 
The distinctions between the three activities are:
\begin{enumerate*}[label=(\arabic*)]
    \item \textit{test automation} only automates the execution of tests and does not include generating test data or oracles;
    \item \textit{test input generation} involves creating test input data, which can be done either manually or automatically, but it does not include test oracle generation;
    \item \textit{test generation} automates the creation of both test inputs and test oracles.
\end{enumerate*}
Among these activities, \textit{test generation} is the least explored, appearing in only three studies, indicating its higher technical challenges compared to the other test activities. 

Other recurring activities include \textit{oracle prediction}, \textit{test execution}, and \textit{attack detection}. These findings highlight a clear focus on minimising manual effort and enabling scalable testing for XR software.

\subsubsection{RQ2.2 Test Concerns} \label{sec:results:test_concern}

Test concerns cover both \emph{test objectives} (e.g., functionality, usability, security) and \emph{test targets} (e.g., cybersickness, collision) of testing.

\paragraph{Test Objective}

\begin{table}
    \rowcolors{1}{white}{gray!20}
    \renewcommand{\arraystretch}{1.2}
    \centering
    \caption{Test objectives in the primary studies}
    \begin{tabularx}{\textwidth}{lXc} \toprule
        \textbf{Test objective} & \textbf{Primary Studies} & \textbf{Number} \\ \midrule
        Functionality & PS3, PS4, PS5, PS7, PS8, PS9, PS12, PS16, PS17, PS22, PS23, PS25, PS26, PS27, PS32, PS33, PS34 & 17 \\
        Usability & PS1, PS7, PS8, PS10, PS11, PS19, PS20, PS21 & 8 \\
        Security & PS14, PS15, PS30 & 3 \\
        Privacy & PS15, PS18 & 2 \\
        Performance & PS7, PS8 & 2 \\
        Load & PS8 & 1 \\
        \bottomrule
    \end{tabularx}
    \label{tab:test_objective}
\end{table}

We categorise the primary studies into six groups based on their test objectives:
\begin{enumerate*}[label=(\arabic*)]
    \item \textit{functionality:} testing whether the functional specifications are correctly implemented,
    \item \textit{usability:} assessing whether SUT negatively impact user experience,
    \item \textit{security:} ensuring the SUT is protected from external attacks,
    \item \textit{privacy:} verifying that user's personal data is safeguarded against local threats,
    \item \textit{performance:} checking whether the SUT meets specific performance requirements (e.g., response time), and
    \item \textit{load:} evaluating the SUT's behaviour, reliability, or stability under stress.
\end{enumerate*}

Table~\ref{tab:test_objective} provides an overview of the test objectives. Among these, \textit{functionality} is the most common objective in the primary studies. It is worth noting that individual studies can cover multiple test objectives.
For instance, PS7~\citep{lehmanARCHIECloudEnabledFramework2023-PS7} addresses \textit{functionality}, \textit{usability}, and \textit{performance}. It proposes a testing framework for system testing of mobile AR apps. The framework provides features such as collecting usability information, including the quality of the user experience in AR scenes; monitoring performance metrics, such as frames per second (FPS) traces, to identify performance dips; detecting functional edge cases through long-term monitoring.

\paragraph{Test Target}

\begin{figure}
    \centering
    \includegraphics[width=1\linewidth]{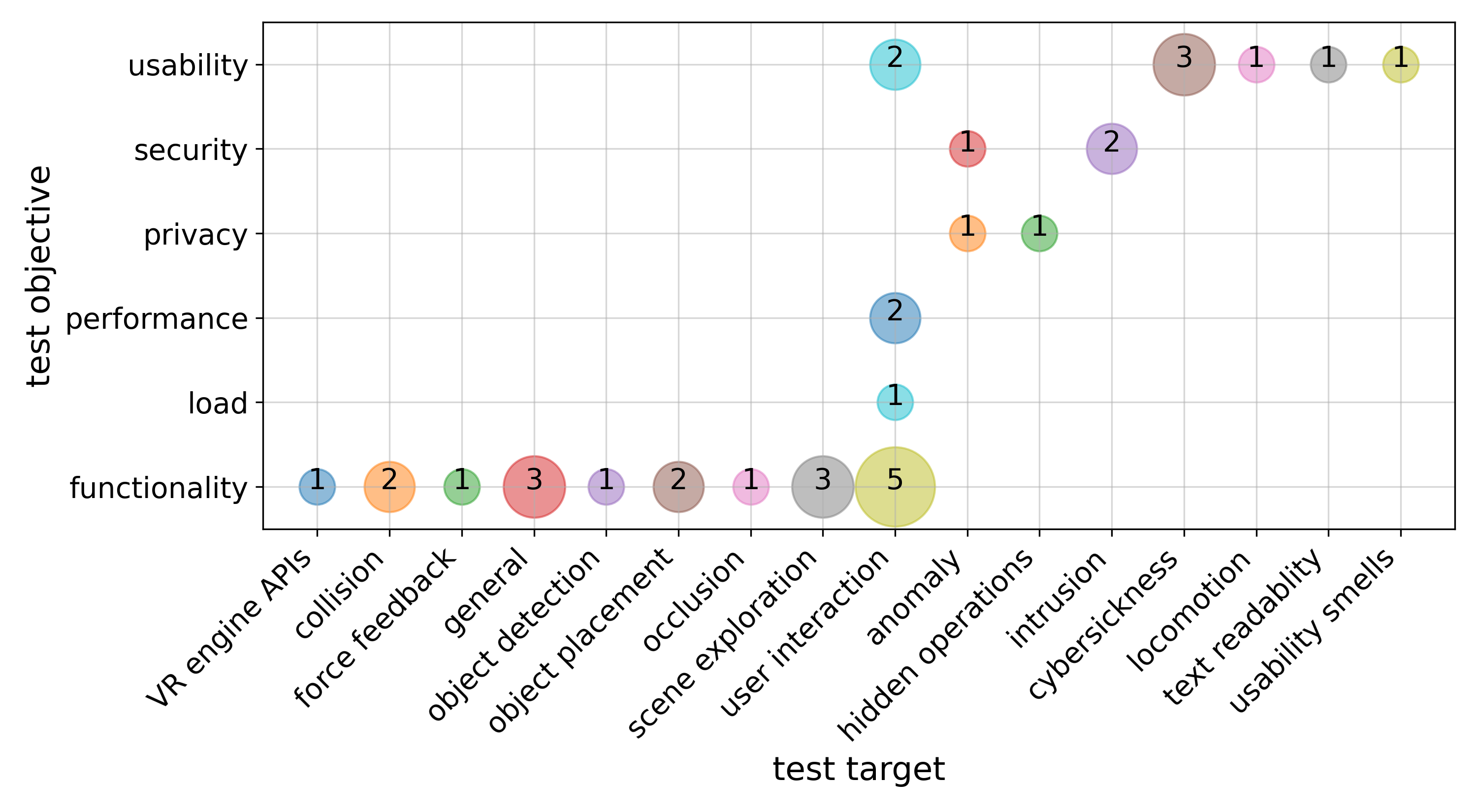}
    \caption{Bubble chart of test objectives against test targets}
    \label{fig:objective_vs_target}
\end{figure}

To illustrate the relationship between test objectives and their associated test targets, we present a bubble chart highlighting how specific test targets align with certain test objectives. Importantly, a single study can address multiple test targets under a single test objective. For example, PS17 focuses on the test objective \textit{functionality} and includes the test targets \textit{collision} and \textit{occlusion}.

Figure~\ref{fig:objective_vs_target} shows the bubble chart. Among the test objectives, \textit{functionality} is the most comprehensive, covering nine distinct test targets. The most studied target under this objective is \textit{user interaction}, which focuses on testing interaction features in XR systems and has been examined in five studies. Other notable targets include \textit{general}, which are the general guidelines for testing XR systems, and \textit{scene exploration}, where testing focuses on exploring the XR scenes, each represented by three studies.

The second most prevalent test objective, \textit{usability}, encompasses five test targets. Among these, \textit{cybersickness} is the most studied, appearing in three studies, while \textit{user interaction} is addressed in two studies.
These findings align with the critical importance of user experience and the operational and interactional aspects of XR systems, emphasising the primary focus on XR software testing efforts.

The other test objectives, i.e., \textit{security}, \textit{privacy}, \textit{performance}, and \textit{load}, cover fewer test targets. Both \textit{security} and \textit{privacy} are linked to two test targets each, while \textit{performance} and \textit{load} are associated with only one test target each.

This analysis underscores the diversity of test targets within each objective and highlights relatively well-explored areas versus those requiring further investigation. 

\subsubsection{RQ2.3 Test Techniques} \label{sec:results:test_technique}

We address this sub-question by analysing the test level (e.g., system testing, unit testing), test type (i.e., black-box or white-box testing) and the specific techniques employed (e.g., random testing, search-based testing).

\paragraph{Test Level}

\begin{table}
    \rowcolors{1}{white}{gray!20}
    \renewcommand{\arraystretch}{1.2}
    \centering
    \caption{Test levels in the primary studies}
    \centering
    \begin{tabularx}{\textwidth}{lXc} \toprule
        \textbf{Test Level} & \textbf{Primary Studies} & \textbf{Number} \\ \midrule
        System Testing & PS1, PS3, PS4, PS5, PS7, PS8, PS9, PS10, PS11, PS12, PS17, PS18, PS19, PS20, PS21, PS22, PS23, PS25, PS26, PS27, PS32, PS33 & 22 \\
        Integration Testing & PS26 & 1 \\
        Unit Testing & PS9, PS26, PS34 & 3 \\
        Non-functional Testing & PS14, PS15, PS30 & 3 \\
        Not Applicable & PS16 & 1 \\
        \bottomrule
    \end{tabularx}
    \label{tab:test_level}
\end{table}

A single study can address multiple test levels. For instance, PS9 introduces preliminary solutions for both unit testing and system-level interaction in VR systems.

Table~\ref{tab:test_level} provides an overview of the test levels explored in the primary studies. \textit{System testing} is the most prominent test level, featured in 22 studies. In contrast, \textit{unit testing} and \textit{non-functional testing} each appear in only three studies, while \textit{integration testing} is represented by a single study.

Test levels do not apply to PS16. The study proposes a technique for recommending potential test types (i.e., animation, colliding, and general) for VR systems but does not explicitly perform any testing actions. Therefore, PS16 is excluded from the data extraction for both \textit{test level} and \textit{test type}.

\paragraph{Test Type}

\begin{table}
    \rowcolors{1}{white}{gray!20}
    \renewcommand{\arraystretch}{1.2}
    \centering
    \caption{Test types in the primary studies}
    \centering
    \begin{tabularx}{\textwidth}{lXc} \toprule
        \textbf{Test Type} & \textbf{Primary Studies} & \textbf{Number} \\ \midrule
        Black-box Testing & PS1, PS11, PS12, PS17, PS18, PS19, PS20, PS21, PS22, PS23, PS25, PS26 & 13 \\
        White-box Testing & PS3, PS4, PS7, PS9, PS10, PS26, PS27, PS32, PS33, PS34 & 10 \\
        Unclear & PS5, PS8 & 2 \\
        Not Applicable & PS14, PS15, PS16, PS30 & 4 \\
        \bottomrule
    \end{tabularx}
    \label{tab:test_type}
\end{table}

We define \textit{white-box testing} as testing that requires access to the source code of the SUT, such as instrumentation or static analysis. In contrast, \textit{black-box testing} does not involve internal information about the SUT. Instead, it relies on analysing the system's input and output behaviour, such as evaluating screenshots or video recordings of specific actions in the XR systems. Notably, a single study may include both black-box and white-box testing methods.

In our analysis, we excluded three studies categorised under the \textit{non-functional testing} level (discussed in the previous paragraph) from the extraction of test types. This is because black-box and white-box testing typically focus on verifying the \textit{functionality} of the SUT. 

Table~\ref{tab:test_type} presents a summary of the findings. The distribution studies utilising \textit{black-box testing} and \textit{white-box testing} are relatively balanced, with 13 and 10 studies, respectively. In addition to the four excluded studies, two studies lack sufficient evidence to determine their test type and are therefore categorised as \textit{unclear}.

\paragraph{Test Techniques}

\begin{table}
    \rowcolors{1}{white}{gray!20}
    \renewcommand{\arraystretch}{1.2}
    \centering
    \caption{Test techniques in the primary studies}
    \centering
    \begin{tabular}{llc} \toprule
        \textbf{Test Technique} & \textbf{Primary Studies} & \textbf{Number} \\ \midrule
        Static Analysis & PS4 & 1 \\
        Dynamic Analysis & PS4 & 1 \\
        Statistical Analysis & PS15 & 1 \\
        Image Processing & PS1 & 1 \\
        Machine Learning & PS11, PS14, PS15, PS17, PS19, PS21, PS23, PS27, PS30 & 9 \\
        Model-based Testing & PS3, PS4, PS5, PS10 & 4 \\
        Search-based Testing & PS32 & 1 \\
        Mutation Testing & PS25 & 1 \\
        Metamorphic Testing & PS17 & 1 \\
        Random Search & PS33 & 1 \\
        Greedy Search & PS33 & 1 \\
        Record and Playback & PS25 & 1 \\
        Runtime Monitoring & PS18 & 1 \\
        Equivalence Partition & PS12 & 1\\
        \bottomrule
    \end{tabular}
    \label{tab:test_technique}
\end{table}

\begin{table}
    \rowcolors{1}{white}{gray!20}
    \renewcommand{\arraystretch}{1.2}
    \centering
    \caption{Evaluation metrics in the primary studies}
    \begin{tabular}{llc} \toprule
        \textbf{Metrics} & \textbf{Primary Studies} & \textbf{Number} \\ \midrule
        Standard ML Metrics & PS11, PS14, PS17, PS18, PS19, PS23, PS27, PS30 & 8 \\ 
        Classification Results &  PS15 & 1 \\
        Manual Validation & PS10, PS12, PS16, PS21 & 4 \\
        Method Coverage & PS32 & 1 \\
        Model Coverage & PS4 & 1 \\
        Object Coverage & PS32, PS33 & 2 \\
        Requirement Coverage & PS5 & 1 \\
        Mutation Score & PS5, PS25 & 2 \\
        SSQ Score & PS21 & 1 \\
        Detected Bugs & PS32 & 1 \\
        Object Detection Success & PS22 & 1 \\
        Suspiciousness Score & PS15 & 1 \\
        \bottomrule
    \end{tabular}
    \label{tab:eval_metric}
\end{table}

Table~\ref{tab:test_technique} summarises the test techniques utilised in the primary studies. A few studies that do not employ specific testing techniques but instead provide general testing guidelines are excluded from the table. Additionally, a single study may apply multiple test techniques. 
In total, 14 distinct techniques are identified, with two being adopted by multiple studies: \textit{machine learning-based testing}, used in nine studies, and \textit{model-based testing}, applied in four studies. These findings align with the characteristics of XR systems. Machine learning techniques are particularly well-suited for handling the rich graphical interfaces of XR systems, such as identifying faults using app screenshots. Meanwhile, model-based testing simplifies the inherent complexity of XR systems by abstracting them into models, facilitating systematic testing.

To further investigate the most predominant test technique, machine learning-based testing, we examine the dataset used in these studies to train the ML models for testing in \S\ref{sec:discussion:dataset_tools}.

Although the remaining test techniques are each represented by only one study, this diversity highlights the successful exploration and adoption of various innovative approaches in the emerging field of XR software testing. Future research could explore the underrepresented areas like integration testing and expand the application of emerging test techniques.

\begin{rqanswer}
\textbf{The answer to RQ2, i.e., the test facets involved in XR applications, is as follows:} 
% \ds{Consider rephrasing this opening sentence to be consistent with the RQ1 summary box.}}

\textbf{Test activities:} The most common test activities involve test automation (11 studies) and test generation (8 studies), reflecting a focus on reducing manual effort in testing XR software.

\textbf{Test concerns:} The primary test objectives are functionality (17 studies) and usability (8 studies). The most frequent test targets for functionality are user interaction (5 studies), while for usability, the key focus is cybersickness (3 studies). These align with critical user experience and interactional aspects of XR systems.

\textbf{Test techniques:} System testing is the dominant test level (22 studies), with black-box and white-box testing nearly balanced. Machine learning is the most prevalent technique (9 studies), followed by model-based testing (4 studies).
\end{rqanswer}

\subsection{RQ3: Evaluation}

We address RQ3 by presenting the evaluation metrics reported in the studies and the evaluation environment used to assess the testing techniques.

Several studies lack concrete evaluation, as they do not fall under the research types of \textit{evaluation research} or \textit{validation research} (\S\ref{sec:research_type}). Specifically, six solution proposals, five empirical studies and two philosophical papers are excluded from the extraction of evaluation metrics and environments.

\subsubsection{Metrics}

Table~\ref{tab:eval_metric} summarises the evaluation metrics used in the primary studies. As discussed in \S\ref{sec:results:test_technique}, ML-based testing is the most prevalent technique, represented by nine studies. Among these, seven studies used \emph{standard ML metrics}, such as \textit{precision}, \textit{recall}, and \textit{F1-score}, to evaluate ML performance. One study relies on \textit{classification results}, which provide general information on the number of correctly classified cases.

The studies identify four kinds of coverage metrics:
\begin{enumerate*}[label=(\arabic*)]
    \item \textit{method coverage:} measures the percentage of methods exercised by tests out of the total number of methods;
    \item \textit{model coverage:} calculates the proportion of states covered by tests in a model as used in model-based testing, such as a finite state machine;
    \item \textit{requirement coverage:} computes the percentage of nodes covered in a requirement flow graph, which is derived from an XR app's scene graph;
    \item \textit{object coverage}, assesses the percentage of interactable objects triggered by tests.
\end{enumerate*}
Coverage metrics are utilised in four studies, with one covering both method and object coverage. Similarly, \textit{manual validation}, where results are manually verified, is also used in four studies.

However, the reliance on manual validation suggests the need for more systematic and automated evaluation frameworks, especially as XR systems grow in complexity.

\subsubsection{Evaluation Environment}

\begin{table}
    \rowcolors{1}{white}{gray!20}
    \renewcommand{\arraystretch}{1.2}
    \centering
    \caption{Evaluation environments in the primary studies}
    \begin{tabular}{llc} \toprule
        \textbf{Evaluation Environment} & \textbf{Primary Studies} & \textbf{Number} \\ \midrule
        HMD & PS10, PS12, PS14, PS21, PS26, PS30 & 6 \\
        Unity Editor & PS16, PS17, PS23, PS26, PS32, PS33 & 6 \\
        Mobile Device & PS18, PS22, PS26, PS27 & 4 \\
        Haptic Device & PS25 & 1 \\
        Cloud & PS15 & 1 \\
        Unclear & PS4, PS5, PS7, PS11, PS34 & 5 \\
        \bottomrule
    \end{tabular}
    \label{tab:eval_env}
\end{table}

To explore what types of environments are involved in the evaluation, such as simulation, mobile devices, and HMDs, we investigate the evaluation environments within the studies. 
Table~\ref{tab:eval_env} provides the results of evaluation environments. The most common evaluation environments are \textit{HMD} (head-mounted display) and \textit{Unity Editor}, each represented in six studies. For clarity, we define the environment \textit{Unity Editor} as the simulation performed within the Unity's \textit{Scene} or \textit{Game} view\footnote{\url{https://docs.unity3d.com/Manual/UsingTheEditor.html}}. \textit{Mobile device} is the second most common environment, used in four studies, while \textit{haptic device} and \textit{cloud} are the least common, each appearing in one study. 

Besides the studies excluded from the extraction of evaluation metrics and environments due to their research types or empirical nature, five studies lack enough details to determine the evaluation environment and are therefore labelled as \textit{unclear}.

The focus on HMDs, Unity Editor, and mobile devices underscores their critical role in real-word testing, while the limited diversity in environments such as cloud-based or haptic devices, suggests opportunities for further exploration and innovation.

Furthermore, we notice some studies utilise existing datasets for evaluation. The details of these datasets (content types, sizes and availability) are discussed in \S\ref{sec:discussion:dataset_tools}.

\begin{rqanswer}
\textbf{The answer to RQ3, i.e., the extent of the testing approaches validated, is as follows:}

Out of 27 studies involving valid test activities, six studies do not provide any evidence on evaluation, leaving 78\% of the studies validated through some form of evaluation.

\textbf{Metric:} The most common evaluation metrics are standard machine learning metrics, for evaluating machine learning-based techniques. Additionally, manual validation and different types of coverage metrics are equally prevalent.

\textbf{Environment:} The most frequently used evaluation environment HMD, Unity Editor, and mobile device, reflecting the typical platforms for XR application development and testing.
\end{rqanswer}

\section{Discussion} \label{sec:discussion}

In this section, we discuss the findings and implications of this mapping study. Specifically, we address
\begin{enumerate*}[label=(\arabic*)]
    \item the key insights and lessons learned from our methodology;
    \item the datasets and tools utilised or proposed in the primary studies;
    \item the implications for practitioners; and
    \item the remaining challenges and future research directions identified through our analysis.
\end{enumerate*}

% This section
% \begin{enumerate*}[label=(\arabic*)]
%     \item investigates the datasets and tools utilised or proposed in the primary studies to facilitate future research, and
%     \item discusses the challenges and future research directions identified in our study.%, incorporating insights from studies excluded during the selection process to propose meaningful research opportunities.
% \end{enumerate*}

\subsection{Mapping Study Methodology}

While conducting this mapping study, we carefully considered methodological choices that could influence our findings. Our approach embraces the diverse nature of XR testing research while acknowledging its potential impacts on interpretation.

While differences between research types or publication venues may yield varying depths of evidence, this diversity enhances the value of our mapping study. By capturing the full spectrum of XR testing research, we provide a more accurate representation of the field's current state.

Following the guidelines by~\citet{petersen2015GuidelinesConductingSystematic}, we deliberately chose an inclusive approach without applying quality assessments during selection. We acknowledge this introduces certain limitations as analysing heterogeneous studies collectively may obscure category-specific characteristics.
Despite potential influences on the interpretation of trends, we believe the benefits of comprehensive coverage outweigh these limitations for mapping the emerging research area of XR testing.

\subsection{Datasets and Tools} \label{sec:discussion:dataset_tools}

To facilitate future research and practices in XR testing, we present an in-depth investigation of the datasets and tools identified in our primary studies.

This subsection examines
\begin{enumerate*}[label=(\arabic*)]
    \item datasets used for training ML models in ML-based testing techniques,
    \item datasets for evaluating testing techniques,
    \item industrial tools employed or referenced in the studies, and
    \item research tools used or proposed within the studies.
\end{enumerate*}
The availability of the datasets and tools is determined as of the submission date of this mapping study (December 2024). The resources are organised and can be accessed at \url{https://sites.google.com/view/xr-testing}.

\subsubsection{Datasets for Training} \label{sec:discussion:dataset_training}

As discussed in \S\ref{sec:results:test_technique}, nine primary studies (PS11, PS14, PS15, PS17, PS19, PS21, PS23, PS27, PS30) utilised ML-based techniques for testing XR apps. To better understand their capabilities and provide valuable resources for future research and practice, we analyse the datasets used for training the ML-based techniques. Among the nine studies, six provide detailed dataset information. We examine their content type, training set size (excluding test sets), data source, and availability, and summarise our findings in Table~\ref{tab:datasets_training}.% summarises the findings from the six studies based on the information reported in the respective works.

\begin{table}
    \rowcolors{1}{white}{gray!20}
    \centering
    \caption{Training datasets used for machine learning-based testing approaches}
    \begin{tabular}{llrlc} \toprule
        \textbf{Study} & \textbf{Content} & \textbf{Size} & \textbf{Source} & \textbf{Avail.} \\ \midrule
        PS11 & Image & \num{600} & experiments & F \\
        PS19 & Image & \num{20000} & Steam & F \\
        PS21 & Video & \num{61} & UCSD Ped1 \& Ped2, Avenue, KITTI & T \\
        PS23 & Image & \num{720} & Unity Mars & T \\
        PS27 & Image & $\sim$~\num{2740} & Google Play \& GitHub & F \\
        PS30 & Traffic \& attacks & $\sim$~\num{848000} & CIC-IDS2017 & T \\
        \bottomrule
    \end{tabular}
    \label{tab:datasets_training}
\end{table}

% We provide a brief overview of each dataset and its specific task for testing XR applications, referring to each dataset as ``Dataset PSx" based on the corresponding primary study.

\textbf{Dataset PS11} consists of 600 images of XR scenes, containing some texts in their background. Each image is labelled whether the text is readable or not by human participants and features various configurations of font styles and background textures. The dataset is not publicly available.

\textbf{Dataset PS19} is a subset of \num{20000} stereoscopic screenshots, randomly sampled from an original training set of \num{154566} screenshots, collected from 288 VR apps on Steam\footnote{\url{https://store.steampowered.com/}}. Steam is one of the largest platforms for video games, including VR apps. % \ds{Just screenshots without labels? Label information would be useful for readers in all datasets.}

\textbf{Dataset PS21} is based on multiple datasets, comprising a total of 61 video clips, each containing 200 frames, to train a model for measuring exceptional motion in VR video content that contributes to cybersickness. The original datasets are UCSD Ped1 and Ped2~\citep{UCSDPed}, Avenue datasets~\citep{AvenueLu}, and KITTI benchmark datasets~\citep{KITTI}, all are publicly available.

\textbf{Dataset PS23} consists of 720 screenshots from a basic AR scene provided by Unity Mars\footnote{\url{https://unity.com/products/unity-mars}}, a Unity extension for AR/MR content development. The dataset is labelled via crowdsourcing to identify realistic object placement. It is used to train a model to identify object misplacement issues in AR systems, capturing variations in placement gaps, distance, and viewing angles.

\textbf{Dataset PS27} includes \num{3043} screenshots from 21 AR apps sourced from the Google Play Store and GitHub. 
With 90\% (approximately \num{2740} screenshots) allocated for training a model to detect object misplacement issues in AR systems. The dataset is labelled via crowdsourcing to provide placement information. However, the exact numbers of screenshots in the training and testing subsets are not specified in the paper, and the dataset is currently not publicly accessible.

\textbf{Dataset PS30} utilised the Intrusion Detection Evaluation Dataset (CIC-IDS2017)\footnote{\url{https://www.unb.ca/cic/datasets/ids-2017.html}}, containing over 2.8 million network traffic instance, including normal traffic and attacks like DoS and DDoS. Reformatted for binary classification (attack vs benign), it comprises \num{1211327} instances, 70\% are used for training. Notable discrepancies in reported sample sizes between subsets, therefore the training set size is (70\% of \num{1211327}, which is approximately \num{848000}) recalculated for consistency.

% PS14 and PS15 do not provide details about the datasets used. PS17 employs a deep reinforcement learning technique to generate test data, in which data are generated through the exploration and perception of an autonomous agent within a virtual environment.

% As ML-based testing techniques often evaluate their performance using the same dataset from which their training subset is derived, these test sets are not discussed in Section~\label{sec:discussion:dataset_training}.

Overall, the prevalence of image-based training datasets highlights the potential of image-based techniques to address a wide range of software testing tasks for XR applications effectively.

\subsubsection{Datasets for Evaluation} 

This section focuses on evaluation datasets, potentially encompassing diverse data points or scenarios, offering broader applicability for testing methodologies, empirical studies, and potential reuse in future research. Isolated research prototypes or limited open-source applications are not considered comprehensive datasets.

As discussed in \S\ref{sec:discussion:dataset_training}, ML-based techniques often evaluate their performance using test datasets, i.e., subsets derived from the same datasets as their training data. Detailed information about these evaluation sets is omitted to avoid redundancy, as they may only differ from the training sets in size. Apart from these, most studies utilised research prototypes or basic open-source applications. % \ds{Don't follow the meaning of this sentence. Describing the reasons for excluding?}

Two empirical studies, PS2 and PS6, present independent datasets for evaluation. \textbf{Dataset PS2} consists of 390 mobile AR apps from the Google Play Store to conduct an empirical study on user privacy concerns in mobile AR apps. However, this dataset is not available.
\textbf{Dataset PS6} collects 368 real bugs from open-source WebXR projects, labelled with their bug symptoms and root causes and is publicly accessible.

We want to know that multiple studies (PS16, PS17, PS31, PS32, PS33) utilised a dataset called \textbf{Unity List}, which is no longer accessible\footnote{According to Unity List's X homepage \url{https://x.com/unitylist}, it is no longer available.}.

\subsubsection{Industrial Tools}

\begin{table}
    \rowcolors{1}{white}{gray!20}
    \centering
    \caption{Industrial tools in the primary studies. OSS indicates if the tool is open-source or not.}
    \begin{tabular}{l|cccc} \toprule
        \textbf{Name} & \textbf{Platform} & \textbf{Input} & \textbf{Test Type} & \textbf{OSS} \\ \midrule
        UTF & Unity & Test scripts & Unit & T \\
        XRI & Unity & Interaction designs & N/A & T \\
        Airtest & Unity, Cocos\footnote{\url{https://www.cocos.com}} & Test scripts & Scene & T \\
        AltUnity Tester & Unity, Unreal & Test scripts & Scene & F \\
        ML-Agents & Unity & Training env. & Scene & T \\
        clumsy & Windows & N/A & Network & T \\
        Wireshark & Windows, Linux, macOS & N/A & Network & T \\
        \bottomrule
    \end{tabular}
    \label{tab:tool_industrial}
\end{table}

Table~\ref{tab:tool_industrial} highlights industrial tools used or referenced in the primary studies. These tools address various testing needs, including GUI, unit, and network testing, as well as one tool for XR interaction development.
For each tool, we outline key details such as supported platforms and engines, input formats, test types, and whether the tool is open-source. This information is intended to guide researchers and practitioners in selecting tools suitable for their testing requirements. 

% Below are brief introductions to each tool:

\textbf{Unity Test Framework (UTF)}\footnote{\url{https://docs.unity3d.com/Packages/com.unity.test-framework@1.1/manual/index.html}} is an official testing tool provided by Unity for unit testing Unity-based projects. It integrates with NUnit~\footnote{\url{https://nunit.org/}}, a unit testing library for .NET languages.

\textbf{XR Interaction Toolkit (XRI)}\footnote{\url{https://docs.unity3d.com/Packages/com.unity.xr.interaction.toolkit@3.0}} is an official Unity package for creating 3D and UI interactions in VR/AR experiences. While it does not directly facilitate XR app testing, it is useful for prototyping research apps that can serve as experimental platforms for testing methodologies.

\textbf{Airtest}\footnote{\url{https://airtest.netease.com/}} is a visual-based UI test automation framework commonly used for video game testing.
It uses screenshot-based locators in test scripts to simulate user actions, making it suitable for dynamic and visually complex interfaces.

\textbf{AltUnity Tester}\footnote{\url{https://alttester.com/tools/}} is a test automation framework designed for games and 3D apps, supporting UI and functional testing. Test scripts interact with Unity elements using identifiers such as object names and tags, simulating user actions.

\textbf{ML-Agents}\footnote{\url{https://github.com/Unity-Technologies/ml-agents}} is an open-source toolkit by Unity for training intelligent agents in Unity-based 2D, 3D, and VR/AR environments using various AI methods. It provides Python APIs for training and Unity C\# scripts for environment simulation. With over 17 example Unity environments, it is well-suited for evaluating XR testing approaches, including agent-based testing~\citep{deandrade2023ExploitingDeepReinforcement-PS17}

\textbf{Clumsy}\footnote{\url{https://jagt.github.io/clumsy/}} and \textbf{Wireshark}\footnote{\url{https://www.wireshark.org/}} are tools for network simulation and analysis. Both were used in PS15 to simulate network- and application-based attacks. These tools are applicable to networked applications, including XR clients and servers, enabling the evaluation of resilience and performance under adverse network conditions.

\subsubsection{Research Tools}

\begin{table}
    \rowcolors{1}{gray!20}{white}
    \centering
    \caption{Research tools in the primary studies}
    \begin{tabularx}{\linewidth}{l|p{2cm}Xp{1.8cm}c}
        \toprule
        \textbf{Name} & \textbf{Source} & \textbf{Function} & \textbf{Platform} & \textbf{Avail.} \\ \midrule
        iv4xr & PS3 & Agent-based testing & N/A & T \\
        ARCHIE & PS7 & Usability testing & Unity & T \\
        MAR-Security & PS18 & Hidden operation detection & Android & T \\
        StereoID & PS19 & Cybersickness detection & N/A & F \\
        PredART & PS23 & Object misplacement prediction & Unity & T \\
        VOPA & PS27 & Object misplacement assessment & N/A & F \\
        % ACL & \citet{yang2016DefectPredictionUnlabeled} & Fault proneness prediction & N/A & T \\ 
        VRGuide & PS32 & VR scene exploration & Unity & T \\
        VRTest & PS33 & VR scene exploration & Unity & T \\
        AutoQuest & \citet{AutoQUEST2013Herbold} & Usability smell detection & N/A & F \\
        TESTAR & \citet{vosTestarScriptlessTesting2021} & Scriptless GUI testing & desktop, web, mobile & T \\
        \bottomrule
    \end{tabularx}
    \label{tab:tool_research}
\end{table}

This section examines research tools specifically designed for XR testing, excluding general tools for tasks like data analysis.
We assess each tool's source (primary studies or external references), key functionalities, supported platforms, and availability.
This analysis is based on publicly available versions, focusing on implementations rather than techniques reported in the papers. While we did not run the tools, we thoroughly reviewed their documentation and repositories. Table~\ref{tab:tool_research} lists the tools analysed.

\textbf{iv4XR}\footnote{\url{https://github.com/iv4xr-project}} is a suite of tools for automated testing for XR applications. It includes frameworks for agent-, model-, and reinforcement learning-based testing, as well as user experience testing.

\textbf{ARCHIE}\footnote{\url{https://github.com/lehmansarahm/ARCHIE}} is a Unity Editor plugin for usability testing in mobile and wearable AR apps. The repository includes Unity-based examples and supports cloud functions.

\textbf{MAR-Security}\footnote{\url{https://github.com/lehmansarahm/MAR-Security}} is a framework for preventing hidden operations in mobile AR apps. Its repository includes an Android project implementing the detection mechanism and scripts for collecting runtime data from Android devices.

\textbf{StereoID}\footnote{\url{https://sites.google.com/view/stereoid}\label{footnote:StereoID}} is a tool for detecting stereoscopic visual inconsistencies linked to cybersickness. However, the tool is not currently accessible. 

\textbf{PredART}\footnote{\url{https://sites.google.com/view/predart2022}\label{footnote:PredART}} includes two types of scripts: a C\# camera control script for Unity projects, and scripts for machine learning model implementation and training.

\textbf{VOPA}\footnote{\url{https://sites.google.com/view/vopa-for-artesting/home}\label{footnote:VOPA}}, is a tool designed to assess virtual object misplacement. However, it is not currently publicly accessible.

\textbf{VRGuide}\footnote{\url{https://sites.google.com/view/vrguide2023}\label{footnote:VRGuide}} and \textbf{VRTest}\footnote{\url{https://sites.google.com/view/vrtest2021}\label{footnote:VRTest}} are automated VR testing tools for scene exploration. While each tool employs different exploration strategies, both provide Unity scripts for their implementations.

\textbf{AutoQUEST}~\citep{AutoQUEST2013Herbold} detects usability smell by analysing recorded user data. While its website\footnote{\url{https://autoquest.informatik.uni-goettingen.de/trac/wiki}} is accessible, not the source code but compiled Java (.jar) files are available.

\textbf{TESTAR}\footnote{\url{https://testar.org/}, \url{https://github.com/TESTARtool/TESTAR_dev}}~\citep{vosTestarScriptlessTesting2021} is an open-source tool for scriptless automated testing of desktop, web and mobile apps at the GUI level. The repository includes documentation for setup and execution. PS24~\citep{pastorricos2022ScriptlessTestingExtended-PS24} references it as a tool that can extend for scriptless testing in XR environments.

\subsection{Implications for Practitioners}

Based on our analysis of datasets and tools referenced in the primary studies, our mapping study reveals two useful insights for XR practitioners.

First, regarding tool selection guidance, Table~\ref{tab:tool_industrial} provides a curated selection of industrial tools organised by platform and testing task. While our findings are based on the primary studies selected for this mapping study, we acknowledge that additional options like Meta XR Simulator (discussed in \S~\ref{sec:intro}) may also be valuable for certain testing scenarios.

Second, concerning research-to-practice opportunities, Table~\ref{tab:tool_research} highlights the research tools that address gaps in current industrial offerings. Though these may require additional implementation effort, they provide cutting-edge capabilities for organisations with specialised testing needs or those seeking competitive advantages in XR quality assurance.

\subsection{Challenges and Future Research Directions}

This section explores the open issues and potential future research directions based on the findings of this mapping study.

During the study selection process (cf. \S\ref{sec:search_selection_strategy}), some studies are excluded as they do not directly align with the focus on testing-related research. However, these studies address challenges that could inspire novel testing approaches by being adapted to specific XR testing needs.
By integrating insights from these excluded studies with the findings from our mapping study, we aim to present meaningful and actionable future research directions to advance XR software testing.

\subsubsection{Interaction Formalisation}

As discussed in \S\ref{sec:results:test_concern}, \textit{user interaction} is the most common testing test objective for functional testing, indicating the importance of interaction testing in XR apps. In \S\ref{sec:results:topics}, we classified \textit{scene testing} studies that validate XR functionality through interactions with virtual objects and scene navigation. However, these approaches provide limited context on the specific interactions required to trigger objects (e.g., touching) or complete navigation tasks (e.g., reaching a destination).

Unlike 2D GUI apps, where interaction types are relatively straightforward, XR apps' 6DOF nature demands more diverse interaction types. Moreover, XR interaction methods may vary based on the deployment platform and device capabilities.

Drawing from prior research, formal gesture descriptions have proven effective in automating UI testing for mobile apps~\citep{hesenius2014AutomatingUITests} and could similarly benefit XR apps. However, this requires a predefined set of XR-specific interaction types, which remains an open challenge~\citep{borsting2022SoftwareEngineeringAugmented}. Leveraging these predefined interactions could support cross-device compatibility testing and facilitate the development of reusable testing frameworks for diverse XR platforms.

We recommend systematic empirical studies to categorise XR-specific interactions (e.g., gestures, haptic feedback) by analysing documentation from XR development platforms and open-source projects to create standardised interaction taxonomies.

\subsubsection{Test Oracle Automation} \label{sec:discussion:test_oracle}

In \S\ref{sec:results:test_activity}, we identified \textit{test automation}, \textit{test input generation}, and \textit{test generation} as the most frequent test activities for XR apps. Among these, \textit{test generation}-- which involves generating both test inputs and oracles--remains the least explored.
Non-crashing functional bugs often require manual validation, with current approaches focusing primarily on crash bugs due to the lack of automated oracles~\citep{suFullyAutomatedFunctional2021}.
Automating oracles is crucial for overcoming this bottleneck and advancing automated testing~\citep{barr2015OracleProblemSoftwarea}. 

While some research has addressed the oracle problem for XR apps, the specific oracles needed to validate functionality remain unclear and vary by system~\citep{pastorricos2022ScriptlessTestingExtended-PS24}. For example, detecting collision and object misplacement may require distinct oracles, each demanding tailored techniques. Addressing this gap necessitates a deeper understanding of the problem and the development of novel solutions.

We propose 
\begin{enumerate*}[label=(\arabic*)]
    \item investigating which XR app characteristics can serve as reliable test oracles, and
    \item determining the most effective oracle types (e.g., assertions, contracts, or metamorphic relationships~\citep{molinaTestOracleAutomation2025}) for different XR testing scenarios.
\end{enumerate*}

\subsubsection{XR-specific Testing}

XR-specific requirements encompass a wide range of test targets, including real-time collision and occlusion, as well as key AR features such as tracking and registration~\citep{VR/AR}. \S\ref{sec:results:topics} identifies \textit{XR-specific testing} as a primary research focus in XR software testing.

Additionally, studies identified during the selection process provide insights into testing these requirements. For example, several studies~\citep{chengKeyIssuesRealtime2021-PS14, wei2012CollisionDetectionAlgorithm-PS24, xu2023VirtualRealityCollision-PS30, jin2021CapsulebasedCollisionDetection-PS40, zhang2014CollisionDetectionTechnology-PS42} propose effective collision testing techniques. However, these studies mainly focus on experimental simulations and have not been applied to specific XR apps. Their methodologies could be adapted to enable systematic collision testing in XR apps, such as instrumenting specific objectives in an XR app to yield collision information.

The unique nature of XR-specific requirements calls for novel testing methodologies not present in other software domains, underscoring the need for tailored approaches and further research.
For effective testing, we recommend first conducting systematic studies to analyse the software manifestation of XR-specific features. This analysis should identify observable behaviours in XR apps and determine which software testing techniques would be most effective for validating these unique characteristics. Such foundational work is essential before developing specialised XR testing methodologies.

\subsubsection{Software-centric Usability Testing}

Cybersickness is the most common usability issue in XR apps, with several studies proposing software-centric techniques for automated detection, as detailed in \S\ref{sec:results:test_concern}.
Additionally, many user-centric studies explore the nature of cybersickness (\S\ref{sec:related:usability}), providing a foundation for developing software-centric detection methods.

Beyond cybersickness, we identified usability-focused user studies during the study selection process that could inform automated testing techniques from a software-centric perspective.
For example, \citet{KIA2023104107} highlighted factors affecting users' muscular loads during AR app interactions, such as interaction error rates and target size. These factors could be formalised into software models to automate the detection of similar usability issues, addressing a broader range of challenges in XR app usability.

To bridge the gap between software- and user-centric approaches, we recommend integrating findings from user studies into automated testing frameworks. This integration would enable the detection of common usability issues without requiring human evaluation, making usability testing more scalable and consistent across XR apps.

\subsubsection{AI for XR Testing}

Advancements in AI, particularly large language models (LLMs) and reinforcement learning techniques, present opportunities to enhance XR app testing.

As discussed in \S\ref{sec:discussion:test_oracle}, test oracle automation remains a significant challenge in XR testing. While crowdsourcing has been shown to effectively address oracle-related tasks~\citep{Pastore13CrowdOracles, rafi2023PredARTAutomaticOracle-PS23}, recent progress in LLMs offers a potential alternative for automating text-based tasks~\citep{Thomas2024SIGIR}, which benefits the generation of human-readable assertions, validating expected outputs, and synthesizing test expectations from natural language specifications.

The oracle problem in XR systems is complex due to their reliance on 3D graphics. However, multimodal LLMs, which process both textual and visual information, have demonstrated capabilities in understanding graphical content, ranging from 2D screenshots to 3D assets~\citep{liu2024visiondrivenautomatedmobilegui, qiu2024largelanguagemodelsunderstand}.
These advancements could enable more robust testing of intricate graphics systems, including XR apps.

Furthermore, LLMs have been effectively used to generate unit tests for Unity-based game development~\citep{paduraru2024UnitTestGeneration}. Given the shared Unity platform, these techniques could potentially be adapted for XR app unit testing, further advancing automation in this domain.

In addition, deep reinforcement learning and imitation learning techniques have demonstrated capabilities to both play (complete specific tasks) and test (explore unknown scenarios) video games~\citep{zheng2019WujiAutomaticOnline}. We suggest leveraging these techniques to tackle the interactive challenges of XR app testing.

\section{Conclusion} \label{sec:conclusion}

This paper presents the methodologies, results, and findings of a systematic mapping study on software testing for XR applications. From an initial pool of \num{1167} studies retrieved from a digital library, we selected 34 relevant studies for in-depth analysis.

We classified these studies and extracted meaningful information to address key research questions regarding the current research status, test facets (including test activities, concerns, and techniques), and evaluation methodologies employed in XR testing.
Additionally, we catalogued datasets and tools referenced in these studies, offering a valuable resource for researchers and practitioners to build upon and advance their work.

The mapping study identifies several open issues and outlines promising future research directions. Our findings highlight the growing importance of XR testing and provide a foundation for advancing methodologies to address its unique challenges.
As XR technology rapidly evolves with new platforms, devices and applications, testing methodologies must not only adapt to support these innovations but also leverage the emerging capabilities they offer. Advanced features and hardware capabilities present both challenges and opportunities for testing. Future testing approaches will need to accommodate the increasing complexity of XR environments and the integration of AI-driven behaviours that characterize next-generation XR systems.

In our future work, we plan to focus on the challenge of interaction formalisation for XR testing. By systematically mapping interactions in XR apps to specific user actions, we aim to develop a comprehensive tool capable of automatically generating user action sequences for executing certain testing tasks. The tool would also maintain traceability of action sequences to facilitate bug analysis and reproduction.

\appendix

\newgeometry{left=2cm, right=2cm}

\section{List of primary studies} \label{appendix:PS}
The list corresponds to the studies prefaced with ``PS'' throughout the paper.

\begingroup
\DefTblrTemplate{firsthead, middlehead,lasthead}{default}{} % <---
\DefTblrTemplate{contfoot-text}{normal}{\scriptsize\textit{Continued on the next page}}
\SetTblrTemplate{contfoot-text}{normal}

\begin{longtblr}{ colspec = {lX}, rowhead=1} 
ID & Title \\ \hline
PS1~\citep{jung2017360degStereoImage-PS1} & 360° Stereo image based VR motion sickness testing system \\ 
PS2~\citep{yang2022StudyUserPrivacy-PS2} & A Study of User Privacy in Android Mobile AR Apps \\
PS3~\citep{prasetya2021AgentbasedArchitectureAIEnhanced-PS3} & An Agent-based Architecture for AI-Enhanced Automated Testing for XR Systems, a Short Paper \\
PS4~\citep{tramontanaApproachModelBased-PS4} & An Approach for Model Based Testing of Augmented Reality Applications \\
PS5~\citep{correasouza2018AutomatedFunctionalTesting-PS5} & An automated functional testing approach for virtual reality applications \\
PS6~\citep{liExploratoryStudyBugs2020-PS6} & An Exploratory Study of Bugs in Extended Reality Applications on the Web \\
PS7~\citep{lehmanARCHIECloudEnabledFramework2023-PS7} & ARCHIE++ : A Cloud-Enabled Framework for Conducting AR System Testing in the Wild \\
PS8~\citep{kirayeva2023AutomatedTestingFunctional-PS8} & Automated Testing of Functional Requirements for Virtual Reality Applications \\
PS9~\citep{bierbaum2003AutomatedTestingVirtual-PS9} & Automated testing of virtual reality application interfaces \\
PS10~\citep{harmsAutomatedUsabilityEvaluation2019-PS10} & Automated Usability Evaluation of Virtual Reality Applications \\
PS11~\citep{leykin2004AutomaticDeterminationText-PS11} & Automatic determination of text readability over textured backgrounds for augmented reality systems \\
PS12~\citep{richardgunawan2023BlackboxTestingVirtual-PS12} & Blackbox Testing on Virtual Reality Gamelan Saron Using Equivalence Partition Method \\
PS13~\citep{kilger2021DetectingPreventingFaked-PS13} & Detecting and Preventing Faked Mixed Reality \\
PS14~\citep{odeleye2021DetectingFramerateorientedCyber-PS14} & Detecting framerate-oriented cyber attacks on user experience in virtual reality \\
PS15~\citep{valluripally2023DetectionSecurityPrivacy-PS15} & Detection of Security and Privacy Attacks Disrupting User Immersive Experience in Virtual Reality Learning Environments \\
PS16~\citep{qin2023DyTRecDynamicTesting-PS16} & DyTRec: A Dynamic Testing Recommendation tool for Unity-based Virtual Reality Software \\
PS17~\citep{deandrade2023ExploitingDeepReinforcement-PS17} & Exploiting deep reinforcement learning and metamorphic testing to automatically test virtual reality applications \\
PS18~\citep{lehman2022HiddenPlainSight-PS18} & Hidden in Plain Sight: Exploring Privacy Risks of Mobile Augmented Reality Applications \\
PS19~\citep{li2024LessCybersicknessPlease-PS19} & Less Cybersickness, Please: Demystifying and Detecting Stereoscopic Visual Inconsistencies in Virtual Reality Apps \\
PS20~\citep{sarupuri2018LUTELocomotionUsability-PS20} & LUTE: A Locomotion Usability Test Environment for Virtual Reality \\
PS21~\citep{kim2017MeasurementExceptionalMotion-PS21} & Measurement of exceptional motion in VR video contents for VR sickness assessment using deep convolutional autoencoder \\
PS22~\citep{sendari2020PerformanceAnalysisAugmented-PS22} & Performance Analysis of Augmented Reality Based on Vuforia Using 3D Marker Detection \\
PS23~\citep{rafi2023PredARTAutomaticOracle-PS23} & PredART: Towards Automatic Oracle Prediction of Object Placements in Augmented Reality Testing \\
PS24~\citep{pastorricos2022ScriptlessTestingExtended-PS24} & Scriptless Testing for Extended Reality Systems \\
PS25~\citep{correa2021SoftwareTestingAutomation-PS25} & Software Testing Automation of VR-Based Systems With Haptic Interfaces \\
PS26~\citep{minor2023TestAutomationAugmented-PS26} & Test automation for augmented reality applications: a development process model and case study \\
PS27~\citep{yang2024AutomaticOraclePrediction-PS27} & Towards Automatic Oracle Prediction for AR Testing: Assessing Virtual Object Placement Quality under Real-World Scenes \\
PS28~\citep{andrade2019SystematicTestingVirtual-PS28} & Towards the Systematic Testing of Virtual Reality Programs \\
PS29~\citep{andradeUnderstandingVRSoftware2020-PS29} & Understanding VR Software Testing Needs from Stakeholders' Points of View \\
PS30~\citep{izuazu2023UnravellingBlackBox-PS30} & Unravelling the Black Box: Enhancing Virtual Reality Network Security with Interpretable Deep Learning-Based Intrusion Detection System \\
PS31~\citep{rzig2023VirtualRealityVR-PS31} & Virtual Reality (VR) Automated Testing in the Wild: A Case Study on Unity-Based VR Applications \\
PS32~\citep{wang2023VRGuideEfficientTesting-PS32} & VRGuide: Efficient Testing of Virtual Reality Scenes via Dynamic Cut Coverage \\
PS33~\citep{wang2022VRTestExtensibleFramework-PS33} & VRTest: An Extensible Framework for Automatic Testing of Virtual Reality Scenes \\
PS34~\citep{figueira2022YoukaiCrossPlatformFramework-PS34} & Youkai: A Cross-Platform Framework for Testing VR/AR Apps \\
\hline
\end{longtblr}

\restoregeometry

\bibliography{references}

\end{document}